\documentclass[12pt]{article}
\usepackage[table]{xcolor}
\usepackage{setspace}
\usepackage{float}
\usepackage{amssymb}
\usepackage{amsthm}
\usepackage{amsmath}
\usepackage{latexsym}
\usepackage{epsfig}
\usepackage{graphicx}
\usepackage{wasysym}
\usepackage{threeparttable}
\usepackage{natbib}
\usepackage{color}
\usepackage{epstopdf}
\usepackage{caption}
\usepackage{subcaption}

\usepackage[breaklinks]{hyperref}

\def\boxit#1{\vbox{\hrule\hbox{\vrule\kern6pt
          \vbox{\kern6pt#1\kern6pt}\kern6pt\vrule}\hrule}}

\def\bse{\begin{eqnarray*}}
\def\ese{\end{eqnarray*}}
\def\be{\begin{eqnarray}}
\def\ee{\end{eqnarray}}
\def\bq{\begin{equation}}
\def\eq{\end{equation}}
\def\bse{\begin{eqnarray*}}
\def\ese{\end{eqnarray*}}

\usepackage{mathptmx}      
\usepackage{bm}
\newcommand{\specialcell}[2][c]{%
  \begin{tabular}[#1]{@{}c@{}}#2\end{tabular}}

 {\begin{list}{}%
         {\setlength{\leftmargin}{#1}}%
         \item[]%
 }
 {\end{list}}

\def\bu{\textbf{u}}
\def\bs{\textbf{s}}

\def\bz{\textbf{Z}}

\def\by{\textbf{Y}}

\def\bfbeta{\bm{\beta}}

\def\bftheta{\bm{\theta}}
\def\bfsig{\bm{\Sigma}}

\def\S{\textbf{S}_{r}}
\def\I{\textbf{I}_{n}}
\def\V{\textbf{V}_{\epsilon}}

\begin{document}
\thispagestyle{empty} \baselineskip=28pt

\begin{center}
{\LARGE{\bf A Comparison of Spatial Predictors when Datasets Could be Very Large}}
\end{center}

\baselineskip=12pt

\vskip 2mm
\begin{center}
Jonathan R. Bradley\footnote{(\baselineskip=10pt to whom correspondence should be addressed) Department of Statistics, University of Missouri, 146 Middlebush Hall, Columbia, MO 65211, bradleyjr@missouri.edu},
Noel Cressie\footnote{National Institute for Applied Statistics Research Australia, University of Wollongong, Australia},
Tao Shi\footnote{Department of Statistics, The Ohio State University}
\end{center}
%
%
%
%
\vskip 4mm

\begin{center}
\large{{\bf Abstract}}
\end{center}
In this article, we review and compare a number of methods of spatial prediction. To demonstrate the breadth of available choices, we consider both traditional and more-recently-introduced spatial predictors. Specifically, in our exposition we review: traditional stationary kriging, smoothing splines, negative-exponential distance-weighting, Fixed Rank Kriging, modified predictive processes, a stochastic partial differential equation approach, and lattice kriging. This comparison is meant to provide a service to practitioners wishing to decide between spatial predictors. Hence, we provide technical material for the unfamiliar, which includes the definition and motivation for each (deterministic and stochastic) spatial predictor. We use a benchmark dataset of $\mathrm{CO}_{2}$ data from NASA's AIRS instrument to address computational efficiencies that include CPU time and memory usage. Furthermore, the predictive performance of each spatial predictor is assessed empirically using a hold-out subset of the AIRS data.
\baselineskip=12pt

%
%
%

\baselineskip=12pt
\par\vfill\noindent
\par\medskip\noindent
\clearpage\pagebreak\newpage \pagenumbering{arabic}
\baselineskip=24pt

\section{Introduction} We are in an era of ``big data,'' where the sizes of available datasets are becoming increasingly larger. For example, consider datasets on earnings from the US Census Bureau's {Longitudinal Employer-Household Dynamics} program, on weather from the {National Oceanic and Atmospheric Administration} (NOAA), and on public health from the {Centers for Disease Control and Prevention} (CDC). In the commercial sector, big data is now available using technology that allows companies to gather (anonymously) information on purchases \citep{commerical}. Pharmaceutical organizations amass large amounts of drug-testing data through combinatorial chemistry, medium-to-high-throughput screening (HTS), and other new technologies \citep{pharma}. Many of these datasets can be very large in size; for example, the National Aeronautics and Space Administration (NASA) collects millions of atmospheric $\mathrm{CO}_{2}$ measurements per month over the globe using the Atmospheric Infrared Sounder (AIRS) instrument on the Terra satellite.

As a result, big data is an important and growing topic in statistics. In the spatial-data setting, there are additional challenges. For example, AIRS $\mathrm{CO}_{2}$ data have global extent, but they are spatially sparse. Additionally, they exhibit complex spatial dependencies that may be nonstationary. Thus, the complexity of ``big spatial data'' has motivated many to propose new statistical methodologies for spatial prediction (e.g., see \citealp[][Ch. 4]{cressie-wikle-book}; \citealp{reviewmethods}, for reviews). In particular, there are methods that use separable covariance functions, tapered covariance matrices, composite likelihoods, and low-dimensional latent Gaussian processes. These methodologies are all motivated by the fact that the Gaussian likelihood is difficult to compute when the dataset is large. Specifically, the Gaussian likelihood involves the computation of an inverse and a determinant of an $n\times n$ covariance matrix, a task that is on the order of $n^{3}$ computations, where $n$ represents the size of the spatial dataset.

Despite the growing number of spatial predictors that are becoming available, there has been no comprehensive comparison between (and among) both traditional and modern spatial predictors. Such a comparison would be highly useful to the more general scientific community. In particular, the GIS community often use spatial interpolation and smoothing (e.g., see \citealp{gis_comparison}), and would benefit from such a comparison. Hence, we shall review the parameterization, the algorithm, and the motivation of seven spatial predictors, also considered by \citet{bradleycs2014} in the context of local spatial predictor selection. We consider three traditional spatial predictors, namely traditional stationary kriging, smoothing splines, and negative-exponential distance-weighting; and we consider four more-recently-introduced spatial predictors, namely Fixed Rank Kriging, one based on modified predictive processes, one based on a stochastic partial differential equation, and lattice kriging. We use a benchmark dataset of $\mathrm{CO}_{2}$ data from NASA's AIRS instrument to empirically compare the predictive performances, computation times, and memory usage of these key spatial predictors.

Kriging based on a stationary covariance function has become a method of spatial prediction covered in standard textbooks
\citep[e.g.,][]{cressie,banerjee-etal-2004,shabenberggotway,cressie-wikle-book} and has a rich history \citep[see][and the references therein]{cressiekrig}.
Since this method of spatial prediction has become a staple, we consider it in our study of AIRS $\mathrm{CO}_{2}$ and call the approach \textit{traditional stationary kriging} (TSK). Another common approach is spatial interpolation using splines, which is obtained by minimizing a penalized-least-squares criterion (e.g., see \citealp{wahba}; \citealp{nychka}). Hence, we also consider \textit{smoothing splines} (SSP) in our comparisons.

However, both TSK and SSP are not ``scalable'' to large datasets; for example, they cannot be computed for the entire AIRS dataset for computational reasons. One simple ad hoc solution to this ``big data'' problem is a spatial predictor based on \textit{negative-exponential distance-weighting} (EDW) \citep[see][p. 371, for a discussion on these types of deterministic methods]{cressie}. Here, a datum's negative log weight is proportional to the Euclidean distance from the prediction location to the datum's location (see Section 2.3 for more details on EDW). 

Although EDW is computationally efficient, we are predominantly interested in spatial predictors that are derived from statistical models and are appropriate for big data. For example, low-rank statistical models provide a computationally efficient way to obtain the optimal kriging predictor and associated measures of error. For this reason, low-rank statistical modeling for spatially referenced data is a popular method in the literature. In the spatial univariate setting, see \citet{johan-2006}, \citet{cressie-tao}, \citet{banerjee}, \citet{johan}, and \citet{kang-cressie-2011}. In the spatio-temporal setting, see \citet{wiklecress_spt}, \citet{wikle2001}, \citet{cressie-shi-kang-2010}, \citet{kang-cressie-shi-2010}, \citet{katzfuss_1,katzfuss2012}, and \citet{bradleymst}. In this article, we focus on two low-rank spatial predictors that have motivated much of this literature: \textit{Fixed Rank Kriging} (FRK), and the \textit{Modified Predictive Process} (MPP) approach.

FRK seeks efficient calculation of the kriging predictor 
in the setting where $n$ is very-large-to-massive.  An advantage of FRK is that the inverse of the covariance matrix can be achieved efficiently using the Sherman-Morrison-Woodbury identity \citep[e.g.,][]{henderson-searle}, allowing 
FRK to be scalable (see Section 2.4 for more details). The approach taken by MPP is similar and starts by first predicting a low-rank random effect called the predictive process. Then, predictions of a latent process are found
by multiplying the prediction of the random effect by a set of basis functions (see Section 2.5 for more details). Some have criticized the use of a low-rank representation of a latent Gaussian process and believe that in many settings much of the variability occurs at high frequencies (see \citealp{lindgren-2011}; \citealp{steinr}, for discussions). However, it should be noted that high-frequency or discontinuous basis functions can address this criticism. 

The remaining two spatial predictors are based on imposing more parametric assumptions on the latent random process. One is based on a \textit{stochastic partial differential equation} (SPD) approach proposed by \citet{lindgren-2011}, and the other is \textit{lattice kriging} (LTK) proposed by \citealp{nychkaLK}. These two methods of prediction achieve computational efficiency by placing structure on the precision matrix of the random-effects vector (see Sections 2.6 - 2.7 for more details). These four recent proposals, coupled with the fact that spatial prediction using big datasets is an important problem, adds additional motivation for our comparison.

In Section 2, we present seven methods of spatial prediction, ranging from the classical to the more recent ones designed to handle very-large-to-massive datasets; both deterministic and stochastic spatial predictors are considered. Details surrounding the predictors are presented systematically, along with the motivation behind each spatial predictor. In Section 3, we apply and compare these predictors using different-sized datasets of mid-tropospheric $\mathrm{CO}_{2}$ measurements. We include the computation time and memory usage of each predictor in the comparison, along with an empirical comparison of predictive performance using a hold-out dataset. A concluding discussion is provided in Section 4.

\section{Seven Spatial Predictors} In this section, we provide details on the spatial predictors considered. They are: traditional stationary kriging (TSK), smoothing splines (SSP), negative-exponential-distance-weighting (EDW), Fixed Rank Kriging (FRK), the modified predictive process approach (MPP), the SPDE approach (SPD), and lattice kriging (LTK). Notice that the spatial predictors could be deterministic or stochastic, and we have chosen several that have been proposed recently to handle big spatial datasets. Details of the seven predictors are set out according to: the parameterization associated with each spatial
predictor; the algorithm used to compute the spatial predictor; and the motivation behind the spatial predictor.

Many of the spatial predictors that we consider can be motivated by a {spatial mixed effects} (SME) model \citep[e.g.,][]{johan-2006,johan}:
\begin{align}
\label{cht4.model}
&\mathrm{Data \hspace{2pt} Model:}\hspace{22pt} Z(\bu) = Y(\bu) + \epsilon(\bu)\\
\label{process:general}
&\mathrm{Process \hspace{2pt} Model:} \hspace{10pt} Y(\bu) = \mu(\bu) +\nu(\bu) + \xi(\bu);\hspace{5pt}\bu \in D,
\end{align}
\noindent
where $\epsilon(\cdot)$ represents measurement error; $\mu(\cdot)$ is a deterministic mean function; $\nu(\cdot)$ models small-scale variation; $\xi(\cdot)$ is a term that captures (often non-smooth) micro-scale variation; and $D\equiv \{\bu_{j}: j = 1,...,N\}\subset \mathbb{R}^{d}$ is a generic finite set of prediction locations. All stochastic components, $\epsilon(\cdot)$, $\nu(\cdot)$, and $\xi(\cdot)$ are assumed mutually independent. A very flexible way to represent $\nu(\cdot)$ is through a basis-function expansion,
\begin{align}\label{basis}
\nu(\bu) = \textbf{S}_{r}(\bu)^{\prime}\bm{\eta};\hspace{5pt} \bu \in D,
\end{align}
\noindent
where $\textbf{S}_{r}(\cdot)$ is an $r$-dimensional vector of spatial basis functions and $\bm{\eta}$ is an $r$-dimensional vector of random coefficients.

The spatial random process $Z(\cdot)$ represents the ``data,'' and it is observed over a subset of the spatial domain of interest $D \subset \mathbb{R}^{d}$; that is, $Z(\cdot)$ is observed at locations in the set $D_{O}\equiv \{\bs_{i}: i = 1,...,n\}\subset D$. The latent process $Y(\cdot)$ is of principal interest, and one wishes to predict it from the data $\{Z(\bs): \bs \in D_{O}\}$. It is assumed that $\epsilon(\cdot)$ is a white-noise Gaussian process with mean zero and known var(\(\epsilon(\cdot)\))= $\sigma_{\epsilon}^{2}V(\cdot)$, where $V(\cdot)>0$ is a known function that captures heteroskedasticity. Note that often variance estimates are obtained from equipment calibration and quality assurance, in which case $\sigma_{\epsilon}^{2}$ can be considered as known. Let $\mu(\cdot) \equiv \textbf{x}(\cdot)^{\prime}\bfbeta$, where $\textbf{x}(\bs)$ is a $p$-dimensional vector of known spatial covariates defined on all $\bs \in D$, and $\bfbeta$ is a $p$-dimensional vector of unknown regression coefficients. 

The low-rank representation of $\nu(\cdot)$ requires further explanation. For $i = 1,...,r$, the $i$-th element of $\textbf{S}_{r}(\cdot)$ is given by the function, $S_{i,r}: D \rightarrow \mathbb{R}$; and the $r$-dimensional random vector $\bm{\eta}$ is specified as a Gaussian process with mean zero and $r\times r$ covariance matrix $\textbf{K}$. Finally, the random process $\xi(\cdot)$ is assumed to be a Gaussian white-noise process with mean zero and variance $\sigma_{\xi}^{2}$.

It will be seen below that the SME model motivates many of the stochastic predictors, although clearly not so for the deterministic predictors. Critically, it is not our intention in this article to fit a single stochastic model given by (\ref{cht4.model}) and (\ref{process:general}); rather, we look at each of the spatial predictors algorithmically, as it acts on the data $\{Z(\bs_{i}): i = 1,...,n\}$. We also consider the ``central'' spatial predictor in each case, recognizing that embellishments may be needed in a particular application. Our goal is to make the review and comparison as straightforward and transparent as possible.

\subsection{Traditional Stationary Kriging (TSK)} 

\noindent
\textbf{Its parameterization:} The statistical model from which TSK is an optimal spatial predictor can be defined hierarchically. The data model is given by (\ref{cht4.model}) with $V(\cdot) \equiv 1$, and $\sigma_{\epsilon}^{2}$ known. The process model is given by,
\begin{equation}\label{cht4.geomodel}
Y(\bu) = \textbf{x}(\bu)^{\prime}\bfbeta + \nu(\bu) + \xi(\bu); \hspace{5pt} \bu \in D,
\end{equation}
\noindent
where $\textbf{x}(\bu)$ is a $p$-dimensional vector of known spatial covariates that describes the large-scale variation, $\nu(\bu)$ represents small-scale variation, and independently $\xi(\bu)$ represents fine-scale variation.

The spatial random process $\nu(\cdot)$ is specified to have mean zero and a second-order stationary covariance function,
\begin{equation}
\mathrm{cov}(\nu(\bu + \textbf{h}), \nu(\bu)) \equiv C(\textbf{h}); \hspace{5pt} \textbf{h}\in \mathbb{R}^{d},
\end{equation}
\noindent
where the function $C(\cdot)$ is positive-definite (e.g., Cressie, 1993, p.68). Specifically, in Section 3, we use the exponential covariance function given by,
\begin{equation}
C(\textbf{h}) = \sigma_{0}^{2}\mathrm{exp}\left( -\frac{||\textbf{h}||}{\theta}\right); \hspace{5pt} \textbf{h}\in \mathbb{R}^{d},
\end{equation}
\noindent
where $\theta > 0$ and $\sigma_{0}^{2}>0$. We organize these unknown parameters into the set $\bftheta^{\mathrm{TSK}} \equiv \{\bfbeta,\theta,\sigma_{0}^{2},\sigma_{\xi}^{2}\}$. \\

\noindent
\textbf{The algorithm:} To compute TSK for a given $\bftheta^{\mathrm{TSK}}$, first construct the $n \times n$ covariance matrix,
		\begin{align}
		\bm{\Sigma}(\bftheta^{\mathrm{TSK}}) & \equiv \left(\mathrm{cov}\left(\nu(\bs_{i}),\nu(\bs_{j})\vert \theta , \sigma_{0}^{2} \right): i,j=1,...,n \right) + \sigma_{\xi}^{2} \I + \sigma_{\epsilon}^{2} \I , \label{cht4.covargsk}
		\end{align}
\noindent
where $\I$ is the $n \times n$ identity matrix. Also construct the $n$-dimensional vector,
	\begin{align}
\mathrm{cov}(\bz,Y(\bu)\vert \bftheta^{\mathrm{TSK}}) & = \mathrm{cov}(\bz,\nu(\bu)\vert \bftheta^{\mathrm{TSK}})+ \sigma_{\xi}^{2} (I(\bu = \bs_{1}),...,I(\bu = \bs_{n}))^{\prime},\label{cht4.covecgsk}
	\end{align}
\noindent
where $I(\cdot)$ represents the indicator function. Then define
	\begin{equation}\label{cht4.blupgeo}
	\hat{Y}(\bu, \bz \vert \bftheta^{\mathrm{TSK}}) \equiv \textbf{x}(\bu)^{\prime}\bm{\beta} + \mathrm{cov}(\bz,Y(\bu)\vert \bftheta^{\mathrm{TSK}})^{\prime}\bm{\Sigma}(\bftheta^{\mathrm{TSK}})^{-1}(\bz - \textbf{X}\bm{\beta}),
	\end{equation}
\noindent
where $\textbf{X} \equiv (\textbf{x}(\bs_{1}),...,\textbf{x}(\bs_{n}))^{\prime}$.

Modifying (\ref{cht4.blupgeo}) to be a function only of the data $\bz$, we substitute in the {ordinary least squares} (OLS) estimate for $\bfbeta$ and {maximum likelihood} (ML) estimates of the covariance parameters where the likelihood assumes mean zero, covariance (\ref{cht4.covargsk}), and it uses the detrended data \citep[e.g.,][p. 239 and pp. 291-292]{cressie}. The estimated parameters are denoted as $\hat{\bftheta}^{\mathrm{TSK}}$. In Section 3, TSK is defined by the predictor,  
\begin{equation}
\hat{Y}^{\mathrm{TSK}}(\bu, \bz) \equiv \hat{Y}(\bu, \bz \vert \hat{\bftheta}^{\mathrm{TSK}});\hspace{2pt} \bu \in D.
\end{equation}
\noindent
To compute $\hat{Y}^{\mathrm{TSK}}$, we use the R-package ``geoR'' version 1.7-4 \citep{diggleGeoR}.\\

\noindent
\textbf{The motivation:} The spatial predictor given by (\ref{cht4.blupgeo}) minimizes the mean squared prediction error,
\begin{align}
\nonumber
E\left((Y(\bu) - \hat{Y}(\bu, \bz))^{2}\vert \bftheta^{\mathrm{TSK}} \right),\label{cht4.mingeo}
\end{align}
\noindent
among the class of linear predictors, $\hat{Y}(\bu, \bz) = \ell + \textbf{k}^{\prime}\bz$ (e.g., Cressie, 1993, Section 3.4.5).\\

\subsection{Smoothing Splines (SSP)}

\noindent
\textbf{Its parameterization:} In our implementation of smoothing splines, there is a single parameter that trades off smoothness with goodness-of-fit, which we denote as $\theta^{\mathrm{SSP}} > 0$. \\

\noindent
\textbf{The algorithm:} The smoothing spline predictor, for a given $\theta^{\mathrm{SSP}}$, is 
\begin{align}
& \hat{\by}(\bu, \bz \vert \theta^{\mathrm{SSP}}) \equiv \textbf{x}(\bu)^{\prime}\hat{\bm{\beta}}^{\mathrm{SSP}} + \textbf{W}(\bu)^{\prime}(\textbf{W} + \theta^{\mathrm{SSP}}\I)^{-1}(\bz - \textbf{X}\hat{\bm{\beta}}^{\mathrm{SSP}}),
\end{align}

\noindent
where $\textbf{x}(\bu)$ is a $p$-dimensional vector of known spatial covariates, $\textbf{X} \equiv (\textbf{x}(\bs_{1}),...,$ $\textbf{x}(\bs_{n}))^{\prime}$ is an $n\times p$ matrix, and
\begin{align}
\nonumber
& \hat{\bfbeta}^{\mathrm{SSP}} \equiv (\textbf{X}^{\prime}(\textbf{W} + \theta^{\mathrm{SSP}}\I)^{-1}\textbf{X})^{-1}\textbf{X}^{\prime}(\textbf{W} + \theta^{\mathrm{SSP}}\I)^{-1}\bz.
\end{align}
\noindent
In our implementation, the $(i, j)$-th entry of $\textbf{W}$, say $W_{ij}$, is obtained from a radial basis function as follows,
\begin{equation}
||\bs_{i} - \bs_{j}||^{2}\mathrm{log}\left(||\bs_{i} - \bs_{j}||\right),
\end{equation}
\noindent
and the $n$-dimensional vector $\textbf{W}(\bu)$ has $i$-th entry $||\bu - \bs_{i}||^{2} \mathrm{log}\left(||\bu - \bs_{i}||\right)$ \citep[e.g.,][p. 31]{wahba}.

The value of $\theta^{\mathrm{SSP}}$ is chosen based on minimizing a leave-one-out cross-validation error \citep[][pp. 47 - 52]{wahba}. Denote this minimized value as $\hat{\theta}_{\mathrm{SSP}}$, and hence SSP is defined by the predictor,
\begin{equation}
\hat{Y}^{\mathrm{SSP}}(\bu, \bz) \equiv \hat{Y}(\bu, \bz \vert \hat{\theta}^{\mathrm{SSP}});\hspace{2pt} \bu \in D,
\end{equation}
\noindent
which is a function only of the data $\bz$. To compute $\hat{Y}^{\mathrm{SSP}}$, we use the Matlab (Version 8.0) function ``griddata.''\\

\noindent
\textbf{The motivation:} The parameter $\theta^{\mathrm{SSP}}$ is used to achieve a balance between goodness-of-fit and degree-of-smoothness of the spatial predictor \citep{wahba}. In $\mathbb{R}^{2}$, the smoothing spline predictor is the function $f(\cdot)$ that minimizes the following penalized sum of squares (Wahba, 1990, p.31; Nychka, 2001),
\begin{equation}
\frac{1}{n}\sum_{i = 1}^{n}(Z(\bs_{i}) - f(\bs_{i}))^{2} + \theta^{\mathrm{SSP}}\int \int \left(\frac{\partial ^{2}f(\bu)}{\partial^{2}u_{1}} + 2\frac{\partial ^{2}f(\bu)}{\partial u_{1} \partial u_{2}} + \frac{\partial ^{2}f(\bu)}{\partial^{2}u_{2}} \right)du_{1} du_{2},
\end{equation}
\noindent
for $\bu = (u_{1}, u_{2})^{\prime}$. Its generalization to $\mathbb{R}^{d}$ for any positive integer $d$, is straightforward.

\subsection{Negative-exponential-distance weighting (EDW)}\textbf{Its parameterization:} There is a single parameter used for controlling the weights in negative-exponential-distance weighting, which we denote as $\theta^{\mathrm{EDW}} > 0$.\\

\noindent
\textbf{The algorithm:} The data are weighted based on their Euclidean distance from the prediction location $\bu$. Let $d_{i}(\bu) \equiv ||\bu - \bs_{i}||$ be the Euclidean distance between $\bu$ and $\bs_{i}$. The negative-exponential-distance-weighting predictor, for a given $\theta^{\mathrm{EDW}}$, is

\begin{equation}
\hat{Y}(\bu, \bz \vert \theta^{\mathrm{EDW}}) \equiv \frac{\sum_{i = 1}^{n} \mathrm{exp}\{-\theta^{\mathrm{EDW}} d_{i}(\bu)\} Z(\bs_{i})}{\sum_{i = 1}^{n}\mathrm{exp}\{-\theta^{\mathrm{EDW}} d_{i}(\bu)\}};\hspace{5pt} \bu \in D.
\end{equation} 
\noindent
The value of $\theta^{\mathrm{EDW}}$ is often prespecified in advance. In this article, we use $\theta^{\mathrm{EDW}} = 1$, although other choices are possible, resulting in more or less smoothness of the predicted surface. Then EDW is defined by the predictor,
\begin{equation}
\hat{Y}^{\mathrm{EDW}}(\bu, \bz) \equiv \hat{Y}(\bu, \bz \vert 1);\hspace{2pt} \bu \in D.
\end{equation}
To compute $\hat{Y}^{\mathrm{EDW}}$, we wrote a simple MATLAB script.

\subsection{Fixed Rank Kriging (FRK)} \textbf{Its parameterization:} The statistical model from which FRK is derived as an optimal spatial predictor, can be defined hierarchically. The data model is given by (\ref{cht4.model}) with both $V(\cdot)$ and $\sigma_{\epsilon}^{2}$ known. The process model is,
\begin{equation}
Y(\bu) = \textbf{x}(\bu)^{\prime}\bfbeta +\textbf{S}_{r}^{\mathrm{BI}}(\bu)^{\prime}\bm{\eta} + \xi(\bu); \hspace{5pt} \bu \in D,
\end{equation}
\noindent
where $\textbf{x}(\bu)$ is a $p$-dimensional vector of known spatial covariates that describes the large-scale variation, $\textbf{S}_{r}^{\mathrm{BI}}(\bu)^{\prime}\bm{\eta}$ represents small-scale variation, and independently $\xi(\bu)$ represents fine-scale variation. The $p$-dimensional vector $\bm{\beta}$, the $r$-dimensional random vector $\bm{\eta}$, and the Gaussian white-noise process $\xi(\cdot)$ are all defined below (\ref{basis}). We organize the unknown parameters into the set $\bftheta^{\mathrm{FRK}} \equiv \{\bfbeta,\textbf{K},\sigma_{\xi}^{2}\}$.

The term $\textbf{S}_{r}^{\mathrm{BI}}(\cdot)$ is an $r$-dimensional vector function of bisquare basis functions \citep[e.g.,][]{johan}, and the value of $r$ is specified to be much smaller than $n$. As will be discussed at the end of this section, specifying $r\ll n$ leads to computational advantages.\\

\noindent
\textbf{The algorithm:} Define the $n \times n$ matrix $\textbf{V}_{\epsilon} \equiv \mathrm{diag}\{V(\bs_{1}),...,V(\bs_{n})\}$ and the $n \times r$ matrix $\textbf{S}_{r}^{\mathrm{BI}} \equiv (\textbf{S}_{r}^{\mathrm{BI}}(\bs_{1}),...,$ $\textbf{S}_{r}^{\mathrm{BI}}(\bs_{n}))^{\prime}$. To compute FRK, for a given $\bm{\theta}^{\mathrm{FRK}}$, first construct the $n \times n$ covariance matrix,
\begin{align}
\nonumber
\bfsig(\bftheta^{\mathrm{FRK}}) & \equiv \mathrm{cov}(\textbf{Z}\vert \bftheta^{\mathrm{FRK}}, \S^{\mathrm{BI}}) =  \S^{\mathrm{BI}} \textbf{K} (\S^{\mathrm{BI}})^{\prime} +   \sigma_{\xi}^{2} \I + \sigma_{\epsilon}^{2} \textbf{V}_{\epsilon} \label{cht4.covarfrk},
\end{align}
\noindent
where $\I$ is the $n \times n$ identity matrix. Also construct the $n$-dimensional vector,
\begin{align}
&\mathrm{cov}(\textbf{Z},Y(\bu)|\bftheta^{\mathrm{FRK}},\S^{\mathrm{BI}}) = \S^{\mathrm{BI}} \textbf{K} \hspace{2pt} \S^{\mathrm{BI}}(\bu) + \sigma_{\xi}^{2} (I(\bu =\emph{\textbf{s}}_{1}),...,I(\bu =\emph{\textbf{s}}_{n}))^{\prime};\hspace{5pt}\bu \in D,
\end{align}
\noindent
where recall that $I(\cdot)$ represents the indicator function. Then define
	\begin{align}
	& \hat{Y}(\bu, \bz \vert \bftheta^{\mathrm{FRK}})\equiv \textbf{x}(\bu)^{\prime}\bm{\beta} + \mathrm{cov}(\textbf{Z},Y(\bu)|\bftheta^{\mathrm{FRK}},\S^{\mathrm{BI}})^{\prime}\bfsig(\bftheta^{\mathrm{FRK}})^{-1}(\bz - \textbf{X}\bm{\beta})\label{cht4.frknoest};\hspace{5pt} \bu \in D,
	\end{align}
\noindent
where $\textbf{X} \equiv (\textbf{x}(\bs_{1}),...,\textbf{x}(\bs_{n}))^{\prime}$.

Modifying (\ref{cht4.frknoest}) to be a function only of the data $\bz$, we substitute in the OLS estimate for $\bfbeta$ and the {expectation maximization} (EM) estimates of the covariance parameters; here the likelihood from which the EM estimates are obtained assumes that the detrended data follow a Gaussian distribution with mean zero and covariance (\ref{cht4.covarfrk}) \citep{katzfuss_jsm}. For a review of the EM algorithm in this setting, see \citet{bradley2011}. The estimated parameters are denoted as $\hat{\bftheta}^{\mathrm{FRK}}$. Then FRK is defined by the predictor,
\begin{equation}\label{cht4.frk}
\hat{Y}^{\mathrm{FRK}}(\bu, \bz) \equiv \hat{Y}(\bu, \bz \vert \hat{\bftheta}^{\mathrm{FRK}}); \hspace{5pt}\bu \in D.
\end{equation}
\noindent
To compute $\hat{Y}^{\mathrm{FRK}}$, we use Matlab code that is available on the website //niasra.uow.edu.au/cei/\\webprojects/UOW175995.html.\\

\noindent
\textbf{The motivation:} The spatial predictor given by (\ref{cht4.frknoest}) minimizes the mean squared prediction error,
\begin{align}
\nonumber
& E\left((Y(\bu) - \hat{Y}(\bu, \bz))^{2}\vert \bftheta^{\mathrm{FRK}} \right),
\end{align}
\noindent
among the class of linear predictors, $\hat{Y}(\bu, \bz) = \ell + \textbf{k}^{\prime}\bz$ \citep{johan}.

The primary motivation for FRK, as described in \citet{johan}, is that \(\bfsig(\bftheta^{\mathrm{FRK}})^{-1}\) can be computed efficiently using the Sherman-Morrison-Woodbury formula \citep[e.g.,][]{henderson-searle}: 
\begin{align}
\nonumber
\bfsig(\bftheta^{\mathrm{FRK}})^{-1} & = (\sigma_{\xi}^{2}\I + \sigma_{\epsilon}^{2}\V)^{-1}- (\sigma_{\xi}^{2}\I + \sigma_{\epsilon}^{2}\V)^{-1} \S^{\mathrm{BI}} \\
&\hspace{5pt} \times \{\textbf{K}^{-1}+ (\S^{\mathrm{BI}})^{\prime}(\sigma_{\xi}^{2}\I + \sigma_{\epsilon}^{2}\V)^{-1}\S^{\mathrm{BI}}\}^{-1}(\S^{\mathrm{BI}})^{\prime}(\sigma_{\xi}^{2}\I + \sigma_{\epsilon}^{2}\V)^{-1}. \label{cht4.SMWF}
\end{align}
\noindent
Equation (\ref{cht4.SMWF}) allows efficient computation of $\bfsig(\bftheta^{\mathrm{FRK}})^{-1}$ in (\ref{cht4.frknoest}), since (\ref{cht4.SMWF}) involves inverses of $r \times r$ matrices and a diagonal $n \times n$ matrix. Specifically, the computation involved with computing the right-hand side of (\ref{cht4.SMWF}) is of order $nr^{2}$, which is linear in $n$ \citep{johan}.

\subsection{Modified Predictive Process Approach (MPP)}

\noindent
\textbf{Its parameterization:} The statistical model from which MPP is derived as an optimal spatial predictor, can be defined hierarchically. The data model is given by (\ref{cht4.model}) with $V(\cdot) \equiv 1$, and $\sigma_{\epsilon}^{2}$ is unknown. The process model is given by,
\begin{equation}\label{cht4.geomodel}
Y(\bu) = \textbf{x}(\bu)^{\prime}\bfbeta + \textbf{S}_{r}^{\mathrm{PP}}(\bu; \kappa, \sigma_{\nu}^{2})^{\prime}\bm{\eta} + \xi(\bu); \hspace{5pt} \bu \in D,
\end{equation}
\noindent
where $\textbf{x}(\bu)$ is a $p$-dimensional vector of known spatial covariates, $\textbf{S}_{r}^{\mathrm{PP}}(\bu; \kappa, \sigma_{\nu}^{2})^{\prime}\bm{\eta}$ represents small-scale variability, both $\kappa$ and $\sigma_{\nu}^{2}$ are unknown parameters, and independently $\xi(\bu)$ represents fine-scale variability. The $p$-dimensional vector $\bm{\beta}$, the $r$-dimensional random vector $\bm{\eta}$, and the Gaussian white-noise process $\xi(\cdot)$ are all defined below (\ref{basis}). 

Let $ \{\bu_{1}^{*},...,\bu_{r}^{*}\}\equiv D^{*} \subset D$ be a set of ($r\ll n$) knots over the spatial domain $D$. The $r$-dimensional random vector $\bm{\eta}$ is taken to be Gaussian with mean zero and covariance matrix $\textbf{K}^{*}$, where $\textbf{K}^{*} \equiv \left\lbrace C(\bu_{i}^{*}, \bu_{j}^{*})\right\rbrace$ and $C(\cdot)$ is the exponential covariance function with scaling parameter $\kappa>0$ and variance $\sigma_{\nu}^{2}$. The term $\textbf{S}_{r}^{\mathrm{PP}}(\cdot; \kappa, \sigma_{\nu}^{2})$ is an $r$-dimensional vector function defined as,
\begin{equation}
\textbf{S}_{r}^{\mathrm{PP}}(\bu; \kappa, \sigma_{\nu}^{2})^{\prime} \equiv \textbf{k}(\bu)^{\prime}\left(\textbf{K}^{*}\right)^{-1},
\end{equation}
where $\textbf{k}(\bu) \equiv \left(C(\bu, \bu_{i}^{*}): i = 1,...,r\right)^{\prime}$ also depends on parameters $\kappa$ and $\sigma_{\nu}^{2}$. 

The original predictive process approach, proposed by \citet{banerjee}, did not include $\xi(\cdot)$, and this leads to a variance of the hidden process that is underestimated. Later \citet{finley} introduced the fine-scale variability term $\xi(\cdot)$ into the model, resulting in the {modified} predictive process approach. They model the spatial random process $\xi(\cdot)$ as a mean zero independent Gaussian process such that $\mathrm{var}(\xi(\bu)) = C(\bu, \bu) - \textbf{k}(\bu)^{\prime}\left(\textbf{K}^{*}\right)^{-1}\textbf{k}(\bu)$. This leads to
\begin{align*}
 &\mathrm{var}\left(Y(\bu)\right) = \mathrm{var}\left(\textbf{S}_{r}^{\mathrm{PP}}(\bu; \kappa, \sigma_{\nu}^{2})^{\prime}\bm{\eta} + \xi(\bu)\right) \\
 &= \textbf{k}(\bu)^{\prime}\left(\textbf{K}^{*}\right)^{-1}\textbf{k}(\bu) + C(\bu, \bu) - \textbf{k}(\bu)^{\prime}\left(\textbf{K}^{*}\right)^{-1}\textbf{k}(\bu) = C(\bu, \bu).
 \end{align*}
 Thus, the variance of the exponential covariance function is preserved. We organize the unknown parameters into the set $\bftheta^{\mathrm{MPP}} \equiv \{\bfbeta,\kappa, \sigma_{\nu}^{2},\sigma_{\epsilon}^{2}\}$.\\

\noindent
\textbf{The algorithm:} Markov Chain Monte Carlo (MCMC) techniques are used for inference on parameters in this setting \citep{banerjee,finley}. The prior distributions are taken as $\sigma_{\nu}^{2}\sim\mathrm{IG}(a_{\eta},b_{\eta})$, $\kappa\sim\mathrm{U}(a_{\kappa},b_{\kappa})$, $\sigma_{\epsilon}^{2}\sim\mathrm{IG}(a_{\epsilon},b_{\epsilon})$, and $\bfbeta$ has a flat prior, where $\sigma_{\eta}^{2}$, $\kappa$, $\sigma_{\epsilon}^{2}$, and $\bfbeta$ are assumed mutually independent, $\mathrm{IG}(a,b)$ represents an inverted gamma distribution with parameters $a$ and $b$, and $\mathrm{U}(a,b)$ represents a uniform distribution with parameters $a$ and $b$. Choices for the hyperparameters depend on the application, but in Section 3 we use the suggestions from \citet{spBayes}, who also give details of the MCMC computations.

A difference between MPP and the other stochastic spatial predictors under consideration is that MPP predicts the process $Z(\cdot)$. Recall that the data model is given by,
\begin{equation}\label{cht4.geomodel2}
Z(\bu) = Y(\bu) + \epsilon(\bu); \hspace{5pt} \bu \in D,
\end{equation}
\noindent
and hence MPP predicts the process with the measurement error included. Consequently, MPP predictions will be exactly equal to the training data at training data locations $\{\bs_{i}\}$, which is an undesirable property when $\sigma_{\epsilon}^{2}>0$. Typically, scientific interest is in $Y(\cdot)$, and $\epsilon(\cdot)$ in (\ref{cht4.geomodel2}) should be filtered out.

The MCMC generates samples $\{Z(\bu)_{1},...,Z(\bu)_{L}\}$ from the posterior distribution of $Z(\bu)$. Then MPP is defined by the predictor,
\begin{equation}\label{cht4.pdp}
\hat{Y}^{\mathrm{MPP}}(\bu, \bz) \equiv \frac{1}{L}\sum_{\ell = 1}^{L}Z(\bu)_{\ell}; \hspace{5pt}\bu \in D.
\end{equation}
\noindent
To compute $\hat{Y}^{\mathrm{MPP}}$, we use the R-package ``spBayes'' \citep{spBayes}. \\

\noindent
\textbf{The motivation:} The spatial predictor given by (\ref{cht4.pdp}) minimizes the mean squared prediction error,
\begin{equation}\label{motiv:mpp}
E(Z(\bu) - \hat{Y}(\bu, \bz))^{2};\hspace{5pt} \bu \in D,
\end{equation}
\noindent
where here the expectation is taken over $\bz$, $Z(\bu)$, and $\bm{\theta}^{\mathrm{MPP}}$. As we noted above, instead of $Y(\bu)$, the scientifically-less-interesting quantity $Z(\bu)$ appears in (\ref{motiv:mpp}).
The primary motivation of this approach is that since $r\ll n$ the Sherman-Woodbury-Morrison formula can be used to compute the precision matrix efficiently, and thus it should be scalable for large spatial datasets.

\subsection{SPDE Approach (SPD)}

\noindent
\textbf{Its parameterization:} The statistical model from which SPD is derived as an optimal spatial predictor can be defined hierarchically. The data model is given by (\ref{cht4.model}) with $V(\cdot) \equiv 1$, and $\sigma_{\epsilon}^{2}$ unknown. The process model is given by,
\begin{equation}\label{spdmodel}
Y(\bu) = \textbf{x}(\bu)^{\prime}\bfbeta + \textbf{S}_{r}^{\mathrm{PL}}(\bu)^{\prime}\bm{\eta}; \hspace{5pt} \bu \in D,
\end{equation}
\noindent
where $\textbf{x}(\bu)$ is a $p$-dimensional vector of known spatial covariates that describes the large-scale variation, $\textbf{S}_{r}^{\mathrm{PL}}(\bu)^{\prime}\bm{\eta}$ represents small-scale variability, and the fine-scale variability term $\xi(\cdot)\equiv 0$. The $p$-dimensional vector $\bm{\beta}$ and the $r (>n)$-dimensional vector $\bm{\eta}$ are defined below (\ref{basis}). Here the term $\textbf{S}_{r}^{\mathrm{PP}}(\cdot)$ is an $r$-dimensional vector function whose elements are piecewise-linear basis functions; and in contrast to FRK and MPP, $r > n$.

On the Euclidean space, define a set of $r$ knots $ \{\bu_{1}^{*},...,\bu_{r}^{*}\}\equiv D^{*}$, which contains the $n$ locations of $D_{O}$; that is, $r>n$. The $r$-dimensional random vector $\bm{\eta}$ is specified to be a mean-zero Gaussian Markov random field defined on $D^{*}$. The precision matrix associated with $\bm{\eta}$ (i.e., $\textbf{K}^{-1}\equiv \mathrm{cov}\left(\bm{\eta}\right)^{-1}$) is based on parameters $\kappa$ and $\sigma_{\nu}^{2}$. The functional form of this precision matrix, and hence the neighborhood structure of the elements in $\bm{\eta}$, is found by solving a stochastic partial differential equation, which we describe below. We organize the unknown parameters into the set $\bftheta^{\mathrm{SPD}} \equiv \{\bfbeta,\textbf{K}^{-1}, \sigma_{\epsilon}^{2}\}$.\\

\noindent
\textbf{The algorithm:} Bayesian inference proceeds without using MCMC; it is based on {Integrated nested Laplacian approximations} (INLA) in this setting \citep{rue, lindgren-2011}. The INLA algorithm is derived from Laplace approximations of integrals of probability density functions. 

First, priors are chosen for $\bftheta^{\mathrm{SPD}}$. As a default in the R-INLA package, $\bfbeta\sim\mathrm{Gau}(\bm{0},\tau_{\beta}^{2}\textbf{I})$, and $\mathrm{log}\left(1/\sigma_{\nu}^{2}\right)$, $\mathrm{log}\left(\sqrt{8}/\kappa\right)$, and $\mathrm{log}\left(\sigma_{\epsilon}^{2}\right)$ are distributed as Log-Gamma.  Further, $\bfbeta$, $\sigma_{\nu}^{2}$, $\kappa$, and $\sigma_{\epsilon}^{2}$ are assumed to be mutually independent. The values of hyperparameters of the prior distribution are chosen heuristically \citep[][personal communication]{inlaPack} based on default settings of the R-INLA package.

Denote the posterior probability density function of $Y(\bu)$ as $\pi(Y(\bu)\vert \bz)$, and the INLA-approximated version is denoted as $\bar{\pi}(Y(\bu)\vert \bz)$ \citep[e.g.,][Section 3]{rue}. Rejection sampling is then used to generate $L$ values $\{Y(\bu)_{1},...,Y(\bu)_{L}\}$ from $\bar{\pi}(Y(\bu)\vert \bz)$. Then SPD is defined by the predictor,
\begin{equation}\label{cht4.spde}
\hat{Y}^{\mathrm{SPD}}(\bu, \bz) \equiv \frac{1}{L}\sum_{\ell = 1}^{L}Y(\bu)_{\ell}; \hspace{5pt}\bu \in D .
\end{equation}
\noindent
To compute $\hat{Y}^{\mathrm{SPD}}$, we use the R-package ``inla'' \citep{rue,inlaPack}. \\

\noindent
\textbf{The motivation:} The spatial predictor given by (\ref{cht4.spde}) minimizes the (approximate) posterior mean squared prediction error,
\begin{equation}
\int (Y(\bu) - \hat{Y}(\bu, \bz))^{2} \bar{\pi}(Y(\bu)\vert \bz) dY(\bu).
\end{equation}

Computational efficiency is obtained through a connection between Gaussian Markov random fields and Gaussian processes that have a Mat\'{e}rn covariance function,
\begin{equation}\label{cht4.matern}
\frac{\sigma_{\nu}^{2}}{\Gamma(\alpha)2^{\alpha - 1}}(\kappa||\textbf{h}||)^{\alpha}K_{\alpha}(\kappa||\textbf{h}||);\hspace{2pt}\textbf{h} \in \mathbb{R}^{d},
\end{equation}
\noindent
where $K_{\alpha}(\cdot)$ is the modified Bessel function of the second kind of order $\alpha >0$. Here, $0 < \alpha <\infty$ is a smoothing parameter, $\kappa>0$ is a scaling parameter, and $\sigma_{\nu}^{2}$ is the variance parameter. 

A random process $\nu(\cdot)$ in $\mathbb{R}^{d}$ with covariance function given by (\ref{cht4.matern}) is a solution to the following stochastic partial differential equation \citep{whittle}:
\begin{equation}\label{cht4.thespde}
(\kappa^{2} - \Delta)^{\zeta/2}\nu(\bu) = W(\bu); \hspace{2pt} \bu \in \mathbb{R}^{d},
\end{equation}
\noindent
where $W(\cdot)$ is a Gaussian white-noise process with mean zero and variance 1, and $\zeta\equiv \alpha + d/2$ is a positive {integer}, $\kappa > 0$, and $\sigma_{\nu}^{2}>0$.  In (\ref{cht4.thespde}), the {Laplacian} $\Delta$ is defined by,
\begin{equation}
\Delta \equiv \sum_{i = 1}^{d}\frac{\partial^{2}}{\partial^{2}u_{i}}.
\end{equation}

The precision matrix associated with $\bm{\eta}$ (i.e., $\textbf{K}^{-1}\equiv \mathrm{cov}\left(\bm{\eta}\right)^{-1}$) is specified to be a GMRF and is found by substituting $\nu(\bu) = \sigma_{\nu}^{2}\textbf{S}_{r}^{\mathrm{PL}}(\bu)^{\prime}\bm{\eta}$ into Equation (\ref{cht4.thespde}) and solving the stochastic partial differential equation. This solution, which is only for $\zeta$ a positive integer, can be found in Section 2.3 of \citet{lindgren-2011}.

\citet{lindgren-2011} extend this modeling approach to handle nonstationarity by letting some of the parameters depend on spatial coordinates; they find the precision matrix associated with the random vector $\bm{\eta}$ that solves the following stochastic partial differential equation,
\begin{equation}
(\kappa^{2}(\bu) - \Delta)^{\zeta/2}\{\sigma_{\nu}^{2}(\bu)\textbf{S}_{r}^{\mathrm{PL}}(\bu)^{\prime}\bm{\eta}\} = W(\bu); \hspace{2pt} \bu \in \mathbb{R}^{d},
\end{equation}
\noindent
where $\zeta \equiv \alpha + d/2$ is a positive integer, $\kappa(\bu) > 0,$ and $\sigma_{\nu}^{2}(\bu)>0$. \citet{lindgren-2011} propose the model,
\begin{equation}
\mathrm{log}\left(\sigma_{\nu}^{2}(\bu)\right) \equiv \underset{i}{\sum}\beta_{i}^{(1)}B_{i}^{(1)}(\bu)
\end{equation}
\noindent
and
\begin{equation}
\mathrm{log}\left(\kappa^{2}(\bu)\right) \equiv \underset{i}{\sum}\beta_{i}^{(2)}B_{i}^{(2)}(\bu),
\end{equation}
\noindent
where $\{B_{i}^{(1)}\}$ and $\{B_{i}^{(2)}\}$ represent two different finite sets of smooth basis functions. 

Finally, in $\mathbb{R}^{2}$, $\textbf{K}^{-1}$ is specified as follows: $\alpha = 1$ and hence $\zeta = 2$, since $d = 2$; $\{B_{i}^{(1)}\}$ is a set of four spherical basis functions of order three; and $\{B_{i}^{(2)}\}$ is a set of seven spherical basis function of order six \citep[e.g.,][]{lindgren-2011}.

\subsection{Lattice Kriging (LTK)}
\noindent
\textbf{Its parameterization:} The statistical model defining lattice kriging can be defined hierarchically. The data model is given by (\ref{cht4.model}) with $V(\cdot) \equiv 1$; and $\sigma_{\epsilon}^{2}$ is assumed known. The process model is given by,
\begin{equation}\label{wend:process}
Y(\bu) = \textbf{x}(\bu)^{\prime}\bm{\beta} + \textbf{S}_{r}^{\mathrm{WL}}(\bu)^{\prime}\bm{\eta}; \hspace{5pt} \bu \in D,
\end{equation}
\noindent
where $\textbf{x}(\bu)$ is a $p$-dimensional vector of known spatial covariates, $\textbf{S}_{r}^{\mathrm{WL}}(\bu)^{\prime}\bm{\eta}$ represents small-scale variability, and the fine-scale variability term $\xi(\cdot)\equiv 0$. The $p$-dimensional vector $\bm{\beta}$ and the $r (>n)$-dimensional vector $\bm{\eta}$ are defined below (\ref{basis}). Here the term $\textbf{S}_{r}^{\mathrm{WL}}(\cdot)$ is an $r$-dimensional vector function whose elements are ``smooth'' Wendland basis functions; notice that $r>n$.

From \citet{nychkaLK}, define a set of $r$ knots $\{\bu_{1}^{*},...,\bu_{r}^{*}\}\equiv D^{*}$ on a regular grid contained in $D$. Then define the $r$-dimensional random vector $\bm{\eta} \equiv \textbf{B}^{-1} \textbf{e}$, where $\textbf{e}$ is an $r$-dimensional Gaussian random vector with mean zero and variance $\sigma_{\eta}^{2}\textbf{I}_{r}$. Note that $\textbf{B}\bm{\eta} = \textbf{e}$, which is the form of a simultaneous autoregressive (SAR) model, and 
\begin{align}
\nonumber
\textbf{B} & \equiv
\begin{pmatrix}
  4+\kappa^{2} & -1 & 0 &  \cdots & 0\\
  -1 & 4+\kappa^{2} & -1  & \cdots & 0\\
    0 & -1 &  & \cdots & 0\\
  \vdots & & \ddots & & \vdots &\\
      \vdots & &  & -1 & 0\\
    \vdots & & -1 & 4+\kappa^{2} & -1\\
  0& \cdots & & -1 & 4+\kappa^{2}
\end{pmatrix},
\end{align}
 for $\kappa\ge 0$. The elements of $\bm{\eta}$ are arbitrarily ordered based on the locations of the knots. We organize the unknown parameters into the set  $\bftheta^{\mathrm{LTK}} \equiv \{\bfbeta, \sigma_{\eta}^{2}, \kappa\}$.\\

\noindent
\textbf{The algorithm:} Define $\S^{\mathrm{WL}} \equiv (\textbf{S}^{\mathrm{WL}}(\bs_{1}),...,\textbf{S}^{\mathrm{WL}}(\bs_{n}))^{\prime}$. To compute LTK, for a given $\bm{\theta}^{\mathrm{LTK}}$, first construct the $n \times n$ covariance matrix,
\begin{align}
\bfsig(\bftheta^{\mathrm{LTK}}) \equiv \mathrm{cov}(\textbf{Z}\vert \textbf{K},\textbf{S}_{r}^{\mathrm{WL}}) =  \S^{\mathrm{WL}} \textbf{K} (\S^{\mathrm{WL}})^{\prime}+ \sigma_{\epsilon}^{2} \I ,
\end{align}
where recall that $\I$ is the $n \times n$ identity matrix and $\textbf{K} \equiv \mathrm{cov}(\bm{\eta})$. Also construct the $n$-dimensional vector,
\begin{align}
\nonumber
\mathrm{cov}(\textbf{Z},Y(\bu)|\bftheta^{\mathrm{LTK}},\S^{\mathrm{WL}}) = \S^{\mathrm{WL}} \textbf{K} \hspace{2pt} \S^{\mathrm{WL}}(\bu) .
\end{align}
\noindent
Then define,
	\begin{align}\label{cht4.lk}
	& \hat{Y}(\bu, \bz \vert \bftheta^{\mathrm{LTK}})\equiv \textbf{x}(\bu)^{\prime}\bm{\beta} + \mathrm{cov}(\textbf{Z},Y(\bu)|\bftheta^{\mathrm{LTK}},\S^{\mathrm{WL}})^{\prime}\bfsig(\bftheta^{\mathrm{LTK}})^{-1}(\bz - \textbf{X}\bm{\beta});\hspace{5pt} \bu \in D,
	\end{align}
\noindent
where \textbf{X} $\equiv (\textbf{x}(\bs_{1}),...,\textbf{x}(\bs_{n}))^{\prime}$.

Modifying (\ref{cht4.lk}) to be a function only of the data $\bz$, we substitute in the maximum likelihood estimate of $\bftheta^{\mathrm{LTK}}$ (denoted $\hat{\bm{\theta}}^{\mathrm{LTK}}$). Then LTK is defined by the predictor,
\begin{equation}
\hat{Y}^{\mathrm{LTK}}(\bu, \bz) \equiv \hat{Y}(\bu, \bz \vert \hat{\bftheta}^{\mathrm{LTK}}); \hspace{5pt}\bu \in D.
\end{equation}
\noindent
To compute $\hat{Y}^{\mathrm{LTK}}$, we use the R package ``LatticeKrig'' \citep{nychkaLK}. \\

\noindent
\textbf{The motivation:} The spatial predictor given by (\ref{cht4.lk}) minimizes the mean squared prediction error,
\begin{align}
& E\left((Y(\bu) - \hat{Y}(\bu, \bz))^{2}\vert \bftheta^{\mathrm{LTK}} \right),
\end{align}
\noindent
among the class of linear predictors, $\hat{Y}(\bu, \bz) = \ell + \textbf{k}^{\prime}\bz$. A numerical motivation for LTK is that \(\bfsig(\bftheta^{\mathrm{LTK}})^{-1}\) can be found using sparse-matrix techniques \citep{nychkaLK}.

\section{A Comparison of the Seven Spatial Predictors: Mid-Tropospheric $\mathrm{CO}_{2}$ Measurements} The Aqua satellite is part of the Earth Observing System (EOS), which is administered by the National Aeronautics and Space Administration (NASA). The Atmospheric Infrard Sounder (AIRS) is an instrument on board the Aqua satellite that retrieves information on atmospheric $\mathrm{CO}_{2}$. Specifically, the
AIRS instrument collects measurements in the form of spectra that are then converted to mid-tropospheric $\mathrm{CO}_{2}$ values in parts per million (ppm) \citep{chahine}. This information on global $\mathrm{CO}_{2}$ has been used to great effect in raising public awareness on greenhouse gases and in determining policy regarding climate change (e.g., see https://www.ipcc.ch/).

The AIRS instrument records data over swaths (or paths) of the Earth's surface (roughly 800 km wide) and extends
from $-60^{\circ}$ to $90^{\circ}$ latitude. Data are collected on a daily cycle from 1:30 pm to 1:30 am. We use AIRS data recorded from February 1 through February 9, 2010. The collected data are then reported at different spatial resolutions. In this article, we analyze AIRS's level-2 $\mathrm{CO}_{2}$ data, which is reported at a 17.6 km by 17.6 km spatial resolution. 

The resulting AIRS $\mathrm{CO}_{2}$ dataset consists of 74,361 total observations. We would like to compare both the predictive performance and the computational performance of each spatial predictor. However, not every predictor can be computed using all 74,361 observations. For example, it is well known that the traditional predictors TSK and SSP cannot handle datasets this large (or larger). Hence, we subset the globe (i.e., $D$) into a study region that contains a smaller number of data points than found in $\{Z(\bs): \bs \in D_{O}\}$. 

In Figure 1, we display Study Region 1, which covers the Midwest US. Here, there is a total of just $71$ observations available, which we separate into two subsets of size $n = 57$ and $m = 14$. The $n$ observations are the ``training'' data (top panel of Figure 1) used to fit the spatial predictors, and the $m$ observations are the ``validation'' data used to assess the predictive performance of each spatial predictor (bottom panel of Figure 1); notice that we reserve roughly 20$\%$ of the data for validation. Our main reason for analyzing this small study region is to compare the predictive performance of \textit{all} seven spatial predictors, which we do in Section 3.1.

Although we are interested in comparing all the predictors, a number of them are designed to handle larger datasets. In particular, EDW, FRK, MPP, SPD, and LTK are relatively straightforward (but non-trivial) predictors that are intended for large spatial datasets. Consider Study Region 2 in Figure 2, which covers the Americas and western Sahara between longitudes $-125^\circ$ to $3^\circ$ and latitudes $-20^\circ$ to $44^\circ$ (this is the same study region used in \citealp{kang-cressie-shi-2010}). There are $n=12,358$ observations used to train each spatial predictor (top panel of Figure 2), and $m=3,090$ observations used for validation (bottom panel of Figure 2). In Section 3.2, we use the data in Figure 2 to compare these five spatial predictors.

Finally, in Section 3.3, we use the entire dataset in Figure 3, which is computationally feasible only for EDW, FRK, SPD, and LTK, but no longer for MPP. There are $n=59,488$ observations used to train each spatial predictor (top panel of Figure 3), and $m=14,873$ observations used for validation (bottom panel of Figure 3). This is by no means an unusually large dataset (with spatially correlated observations) that one might process; for example, \citet{aritrajsm} and \citet{bradleymst} process datasets on the order of millions.

The {training (validation) data} are referenced by their locations, $D^{\mathrm{trn}}\equiv \{\bs_{j}^{\mathrm{trn}}: j = 1,..., n\}$ ($D^{\mathrm{val}}\equiv \{\bs_{j}^{\mathrm{val}}: j = 1,..., m\}$), where $D_{O} = D^{\mathrm{trn}} \cup D^{\mathrm{val}}$ and $D^{\mathrm{trn}} \cap D^{\mathrm{val}}  = \emptyset$. Hence, the total size of the dataset is $n + m$. We use the validation datasets to assess the predictive performance of each spatial predictor. Define the root average squared testing error (RSTE) associated with the predictor $\hat{Y}^{\mathrm{PRD}}$ as,
\begin{equation}
\label{rste}
\mathrm{RSTE}(\hat{Y}^{\mathrm{PRD}},m) \equiv \left(\frac{1}{m}{\sum}_{j = 1}^{m} (Z(\bs_{j}^{\mathrm{val}}) - \hat{Y}^{\mathrm{PRD}}(\bs_{j}^{\mathrm{val}}, \bz))^{2}\right)^{1/2},
\end{equation}
\noindent
 where ``PRD'' notates a generic predictor. The RSTE will be used to compare each spatial predictor (small values are desirable), PRD = TSK, SSP, EDW, FRK, MPP, SPD, and LTK. 
 
 Another criterion that we consider is the predictive model choice criterion (PMCC) from \citet[][see Equation (27)]{rafteryscore},
 \begin{equation}
 \label{rste}
 \mathrm{PMCC}(\hat{Y}^{\mathrm{PRD}},m) \equiv \frac{1}{m}{\sum}_{j = 1}^{m} \frac{(Z(\bs_{j}^{\mathrm{val}}) - \hat{Y}^{\mathrm{PRD}}(\bs_{j}^{\mathrm{val}}, \bz))^{2}}{\hat{\sigma}^{\mathrm{PRD}}(\bs_{j}^{\mathrm{val}}, \bz)^{2}} - \mathrm{log}\left(\hat{\sigma}^{\mathrm{PRD}}(\bs_{j}^{\mathrm{val}}, \bz)^{2}\right),
 \end{equation}
 \noindent
 where $\hat{\sigma}^{\mathrm{PRD}}(\hspace{2pt}\cdot\hspace{2pt}, \hspace{2pt}\cdot\hspace{2pt})^{2}$ is the model-based posterior variance, and hence we can only compute the PMCC for PRD = TSK, FRK, MPP, SPD, and LTK. Notice that for the SME model in (\ref{cht4.model}) and PRD = TSK, FRK, and LTK,
  \begin{align}
  \nonumber
  &\hat{\sigma}^{\mathrm{PRD}}(\bs, \bz)^{2}\\
  &=\mathrm{var}(\nu(\bs)|\hat{\bm{\theta}}^{\mathrm{PRD}}) + \mathrm{var}(\xi(\bs)|\hat{\bm{\theta}}^{\mathrm{PRD}})
    		 		  -\mathrm{cov}(\textbf{Z},Y(\bs)|\hat{\bm{\theta}}^{\mathrm{PRD}}) ^{\prime}\mathrm{cov}(\textbf{Z}|\hat{\bm{\theta}}^{\mathrm{PRD}})^{-1}\mathrm{cov}(\textbf{Z},Y(\bs)|\hat{\bm{\theta}}^{\mathrm{PRD}}).
  \end{align}
  \noindent
The posterior variance for the predictors that are derived using a fully Bayesian approach are estimated by
  \begin{align}
  \nonumber
  &\hat{\sigma}^{\mathrm{PRD}}(\bs, \bz)^{2}=\left\{
  	\begin{array}{ll}
  		\mathrm{var}\left(Z(\bs)_{\ell}: \ell = 1,...,L\right)  &  \mbox{if } \mathrm{PRD}=\mathrm{MPP},\\
  		\mathrm{var}\left(Y(\bs)_{\ell}: \ell = 1,...,L\right)  &  \mbox{if } \mathrm{PRD}=\mathrm{SPD};\hspace{5pt}\bs \in D^{\mathrm{val}},
  	\end{array}
  \right.
  \end{align}
 \noindent
 where recall $\{Z(\cdot)_{\ell}\}$ and $\{Y(\cdot)_{\ell}\}$ are samples from their respective posterior distributions defined in Sections 2.5 and 2.6, respectively. The PMCC is useful for comparing predictors (small values are desirable) because it incorporates information on the implicit model-based prediction errors. However, it has the limitation of not allowing a comparison to deterministic predictors, which can be done with RSTE.
 
 We are interested in evaluating other properties of the predictors in addition to its predictive performance. In particular, to assess the amount of smoothness in PRD, consider the lag-1 semivariogram,
 \begin{equation}\label{semivariogram}
 {\frac{1}{2|C(1)|}\sum_{C(1)}(\hat{Y}^{\mathrm{PRD}}(\bu_{i}) -\hat{Y}^{\mathrm{PRD}}(\bu_{j})) ^{2},}
 \end{equation}
 
 \noindent
 {where PRD = TSK, SSP, EDW, FRK, MPP, SPD, and LTK,  $C(h) \equiv \{(i,j) : ||\bu_{i} - \bu_{j} || = h\}$, $|C(h)|$ denotes the number of distinct elements in the set $C(h)$, $h$ denotes the spatial lag, and $h = 1$ is in a unit of distance defined by the smallest lag at which a semivariogram can be computed. In Study Regions 1, 2, and 3 the unit of distance is $1.41^{\circ}$, $1.5^{\circ}$, and $1.5^{\circ}$, respectively. A large (small) lag-1 semivariogram in (\ref{semivariogram}) suggests that the map of PRD is non-smooth (smooth). 
 
 The exact specifications of each of the seven spatial predictors can be found in Section 2. Here the covariates are $\textbf{x}((\mathrm{latitude},\mathrm{longitude})^{\prime}) \equiv (1, \mathrm{latitude})$, since it is well known that mid-tropospheric $\mathrm{CO}_{2}$ values display a latitudinal gradient \citep{dorit}; that is, there are $p=2$ covariates. Additionally, the measurement-error variances are assumed known for TSK, FRK, and LTK; in practice, these variances are estimated using a variogram-extrapolation technique used by \citet{kang-cressie-shi-2010} and \citet{katzfuss2012}. We use their estimate of $\sigma_{\epsilon}^{2}$ = 5.6062 $\mathrm{ppm}^{2}$ and, for simplicity, we shall take $V(\cdot) \equiv 1$.

In Sections 3.1 through 3.3, all of our computations are performed on a Dell Optiplex 7010 Desktop Computer with a quad-Core 3.40 GHz processor and 8 Gbytes of memory. It is important to note that the timing and memory-usage results may be different for different machines; however, to illustrate what someone might expect in practice, we use a computer that has the specification of a ``typical personal desktop.''

\subsection{Comparison using a Small Dataset of Mid-Tropospheric $\mathrm{CO}_{2}$}

In this section, we use the data in Study Region 1 displayed in Figure 1, which we process using all seven spatial predictors, namely $\hat{Y}^{\mathrm{TSK}}$, $\hat{Y}^{\mathrm{SSP}}$, $\hat{Y}^{\mathrm{EDW}}$, $\hat{Y}^{\mathrm{FRK}}$, $\hat{Y}^{\mathrm{MPP}}$, $\hat{Y}^{\mathrm{SPD}}$, and $\hat{Y}^{\mathrm{LTK}}$. Maps of the seven spatial predictors are given in Figure 4.

Each spatial predictor displays similar general patterns, with lower $\mathrm{CO}_{2}$ values near the Great Lakes. In general, we can separate the predictors in Figure 4 into two categories: smooth and non-smooth. The two deterministic predictors (SSP and EDW) appear non-smooth, whereas the stochastic spatial predictors appear quite smooth; this is also seen in the lag-1 semivariograms in Table 1. This may be because the stochastic predictors can rely on an underlying stationary process in this setting, where the dataset is small and fairly sparse over the prediction region $D$.

The RSTE results for this example (given in Table 1) indicate that FRK is the individual predictor that appears to have the highest predictive performance, while LTK has the least-favorable predictive performance among the seven spatial predictors; however, it should be noted that the RSTE values are fairly similar across different choices of PRD. The PMCC results for this example (given in Table 1) indicate that TSK, MPP, and FRK appear to have the highest predictive performance, while SPD and LTK have the least-favorable predictive performance among the five stochastic spatial predictors. As expected, there are no difficulties with CPU time and memory usage for this small dataset, and each of the seven spatial predictors were computed in a matter of seconds.

\subsection{Comparison using a Large Dataset of Mid-Tropospheric $\mathrm{CO}_{2}$} 

 It is well known that the inversion of a large $n\times n$ matrix makes TSK and SSP computationally impractical. Hence, for this large dataset in Study Region 2 (see Figure 2) we consider the five spatial predictors that can be computed, namely EDW, FRK, MPP, LTK, and SPD.

Maps of the five spatial predictors $\hat{Y}^{\mathrm{EDW}}$, $\hat{Y}^{\mathrm{FRK}}$, $\hat{Y}^{\mathrm{MPP}}$, $\hat{Y}^{\mathrm{SPD}}$, and $\hat{Y}^{\mathrm{LTK}}$ are given in Figure 5. Each spatial predictor displays similar general patterns; however, in contrast to the results in Section 3.1, the large dataset used in this section shows clearly that MPP is the smoothest predictor, EDW is the least smooth, and FRK, LTK, and SPD have similar patterns of smoothness. These results are further corroborated by inspecting the lag-1 semivariograms in Table 2.

The RSTE results for this example (see Table 2) are fairly constant across different choices of PRD, with MPP (EDW) having the highest (least-favorable) predictive performance as measured by RSTE; recall that MPP is the smoothest spatial predictor. Similar conclusions can be made from the PMCC in Table 2, which indicates that the reduced-rank prediction methods appear to have the highest predictive performances, whereas the full-rank prediction methods appear to have less-favorable predictive performances. The CPU time and memory usage are manageable except for MPP, which has a CPU time of approximately 3.5 hours.

\subsection{Comparison using a Very Large Dataset of Mid-Tropospheric $\mathrm{CO}_{2}$}

 In this section, we use the data in Study Region 3 (the {entire dataset}) displayed in Figure 3, and the four spatial predictors that can process a dataset of this size; that is, we compare EDW, FRK, SPD, and LTK. Note that the MPP predictor is computed using a Metropolis-within-Gibbs sampler, making it too computationally intensive for very large spatial datasets. Coincidentally, the four spatial predictors that can handle datasets of this size do not use MCMC algorithms for statistical inference. Specifically, FRK and LTK are empirical Bayesian, SPD uses a fully Bayesian approach based on \citet{rue}'s INLA algorithm, and EDW does not use a statistical model for inference.

Maps of the four spatial predictors, $\hat{Y}^{\mathrm{EDW}}$, $\hat{Y}^{\mathrm{FRK}}$, $\hat{Y}^{\mathrm{SPD}}$, and $\hat{Y}^{\mathrm{LTK}}$ are given in Figure 6. Similar to the results in Sections 3.1 and 3.2, each prediction method displays very similar patterns. The lag-1 semivariograms indicate that SPD is now the least smooth among the four predictors; LTK retains its property of being much smoother than FRK, EDW, and SPD.

The RSTE results for this example (see Table 3) are fairly constant across different choices of PRD (similar to the results in Sections 3.1 and 3.2), with FRK (EDW) having the highest (least-favorable) predictive performance as measured by RSTE. As in Sections 3.1 and 3.2, PMCC indicates that the reduced-rank method FRK has higher predictive performance than the full-rank methods, SPD and LTK. The CPU time for both FRK and SPD indicate that both of these methods are highly computationally efficient for spatial prediction. Moreover, the memory usage for each predictor is modest. However, EDW and LTK require a significant wait-time to obtain spatial predictions (around 1.5 hours).

\section{Discussion} In this article, we present a comparison of spatial predictors from an algorithmic viewpoint. In particular, we systematically layout the parameterization, the algorithm, and the motivation of three traditional methods of spatial prediction and four more-recently-introduced spatial predictors. The traditional spatial predictors include: traditional stationary kriging (TSK), smoothing splines (SSP), and negative-exponential distance-weighting (EDW). The more-recently-introduced spatial predictors include: Fixed Rank Kriging (FRK), a modified predictive processes approach (MPP), a stochastic partial differential equation approach (SPD), and lattice kriging (LTK). Additionally, we use a benchmark of small, large, and very large mid-tropospheric $\mathrm{CO}_{2}$ datasets to compare computation time, memory-usage, and the prediction performance of each spatial predictor.

Recent advances in technology, such as remote sensing, have made large-to-massive spatial datasets more available, making spatial prediction using big datasets an important and growing problem in the statistics literature. Consequently, the algorithmic concerns of CPU time and memory-usage are featured in our comparison along with predictive performance. 

Of the seven predictors we consider, FRK and SPD perform extremely well in terms of CPU time and memory-usage. However, the remaining five spatial predictors are not as efficient. Both EDW and LTK were scalable to the very large benchmark dataset, but the wait-time was rather large (approximately 1.5 hours for each). It is well known that TSK and SSP have very poor CPU time and memory-usage properties for large datasets and, hence, we were only able to use these predictors using the small benchmark dataset. The MPP predictor also has limitations in CPU time; consequently, we were only able to use MPP on the small and large benchmark datasets, the latter dataset resulting in a significant wait-time (around 3.5 hours).

When visually comparing each of the seven spatial predictors, we see that they each display roughly the same general pattern. From an algorithmic point-of-view, this is to be expected, since if the signal-to-noise ratio is ``large enough,'' then any local-averaging scheme should be able to pick large-scale patterns. These visual patterns are further corroborated using the lag-1 semivariogram, which is consistently smaller (larger) for MPP (EDW and SPD). Of the three predictors for which PMCC can be computed for all study regions, FRK has much higher predictive performance than SPD and LTK. The other predictor that can be computed for all three study regions, EDW, had least-favorable predictive performance (among FRK, SPD, LTK and EDW) according to the RSTE criterion. 

Section 3 has filled a need for empirical comparisons between reduced-rank and full-rank spatial predictors; the results shed light on the recent criticisms of reduced-rank statistical modeling \citep{lindgren-2011,steinr} despite the fact that it has been shown to do well in other settings \citep[see, e.g.,][]{wiklecress_spt,johan-2006, cressie-tao,banerjee, johan,finley,katzfuss_1, cressie-shi-kang-2010,kang-cressie-shi-2010,kang-cressie-2011,katzfuss2012}. In terms of predictive performance as measured by RSTE and PMCC, our results on a benchmark dataset of $\mathrm{CO}_{2}$ data from NASA's AIRS instrument showed that reduced-rank methods outperform the viable full-rank alternatives.

\section*{Acknowledgments} This research was partially supported by the U.S. National Science Foundation (NSF) and the U.S. Census Bureau under NSF grant SES$\--$1132031, funded through the NSF-Census Research Network (NCRN) program. It was also partially supported by NASA's Earth Science Technology Office through
its Advanced Information Systems Technology program Grant NNH08ZDA001N. This research is partially supported by NSF grant DMS-1007060 and DMS-1308458.

\bibliographystyle{jasa}
\bibliography{myref}

\newpage
\section*{Figures and Tables}
\begin{figure}[H]
\begin{center}
\begin{tabular}{c}
\includegraphics[width=8cm,height=8cm]{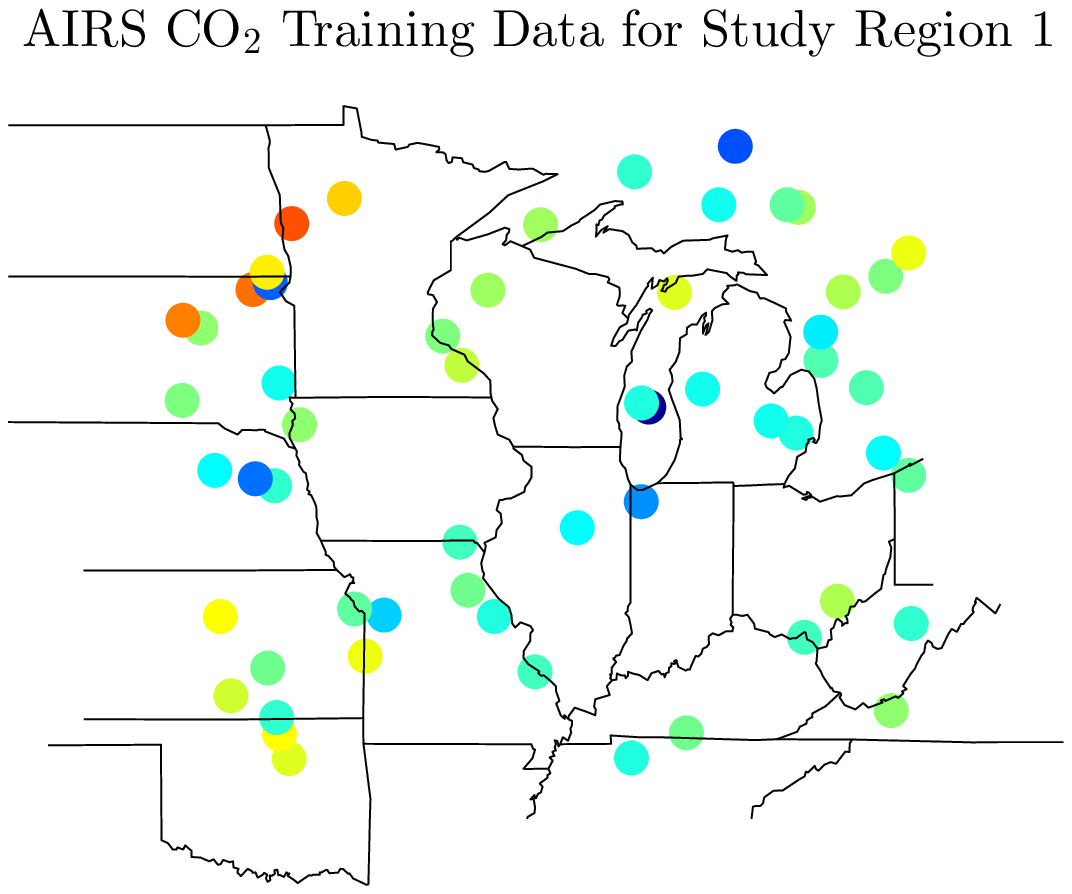}\\
\newline
\includegraphics[width=9.5cm,height=9.5cm]{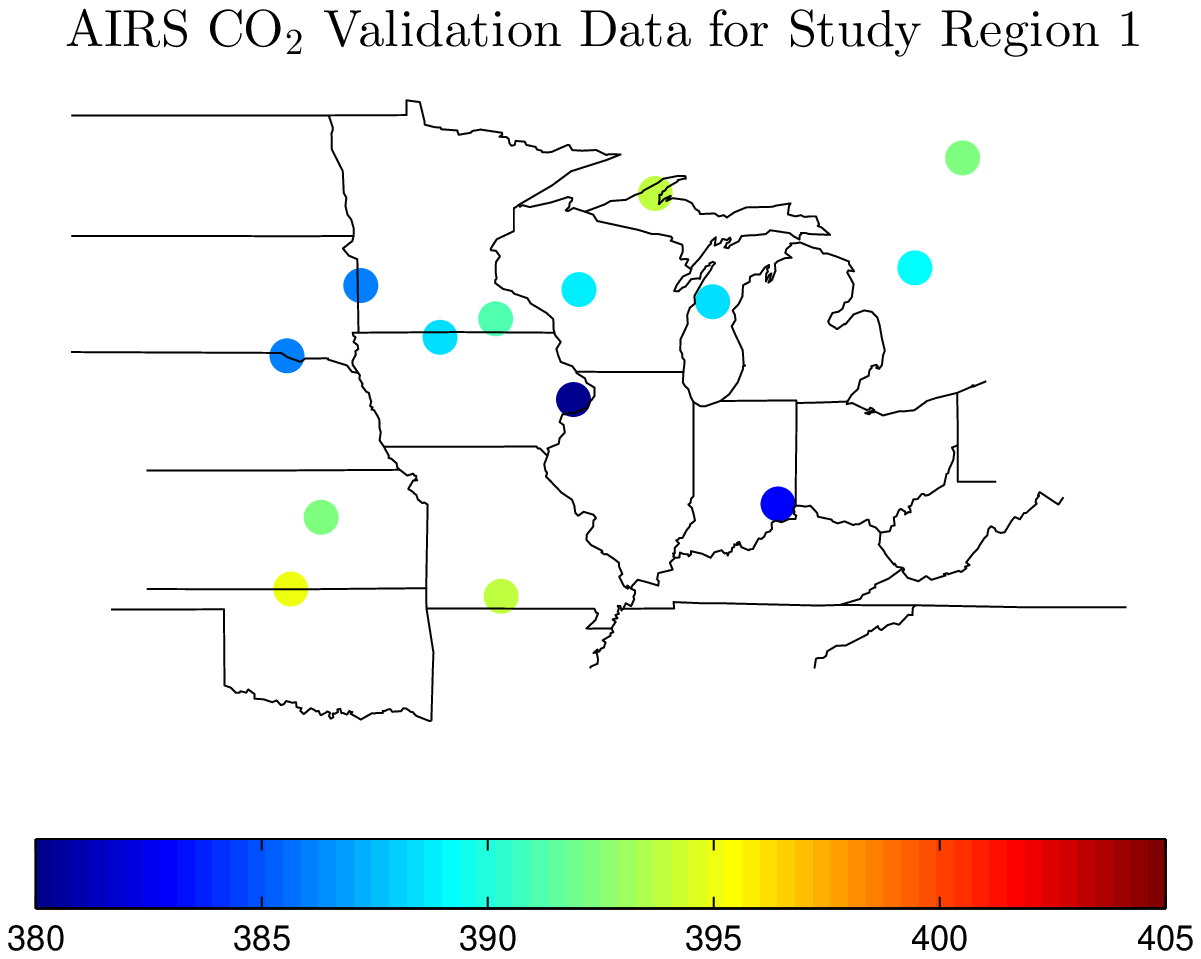}\\
\end{tabular}
\caption{A spatial dataset made up of 9 days of measurements of mid-tropospheric $\mathrm{CO}_{2}$ in parts per million (ppm). The data considered are between $-49^{\circ}$ degrees and $36^{\circ}$ degrees latitude and $-80^{\circ}$ degrees and $-99.5^{\circ}$ degrees longitude, from Februrary 1 through Februrary 9, 2010. The data are randomly split into observed and testing datasets with $n = 57$ and $m = 14$, respectively. }\label{fig:6}
\end{center}
\end{figure}

\clearpage
\begin{figure}
\begin{center}
\begin{tabular}{c}
\includegraphics[width=8.5cm,height=8.5cm]{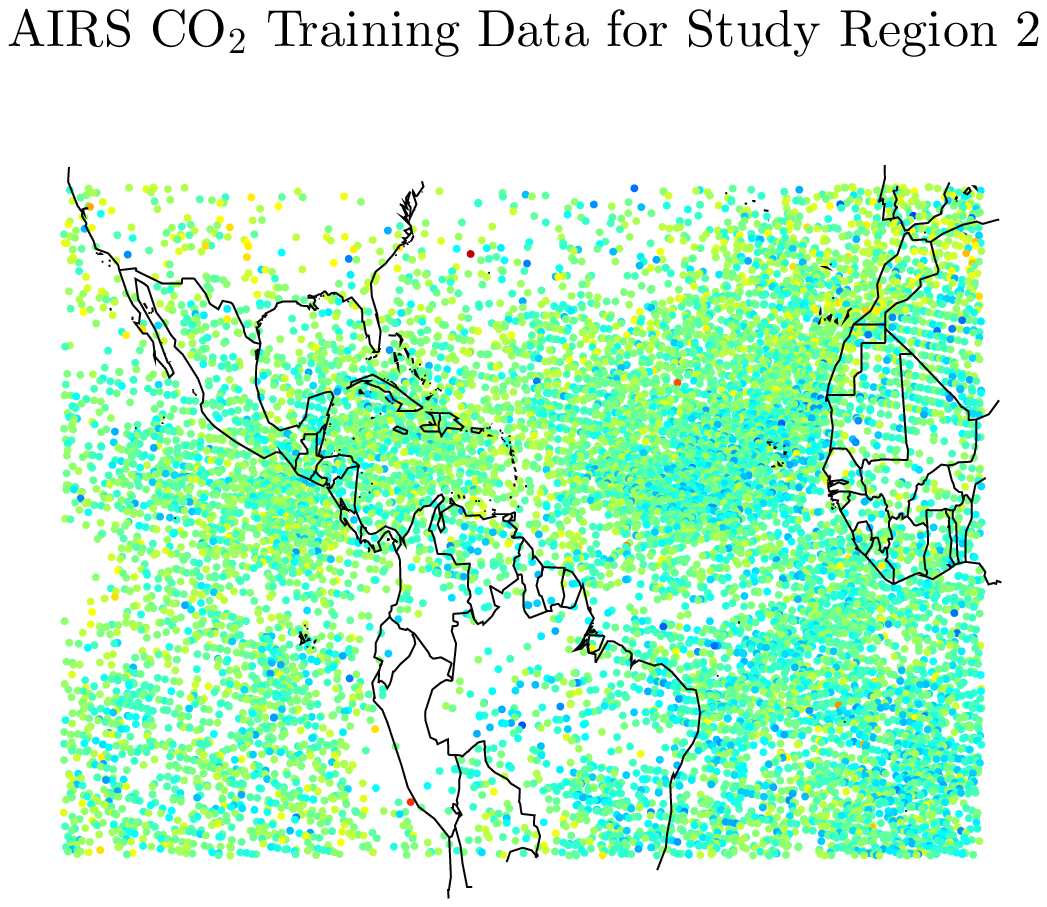}\\
\newline
\includegraphics[width=10cm,height=10cm]{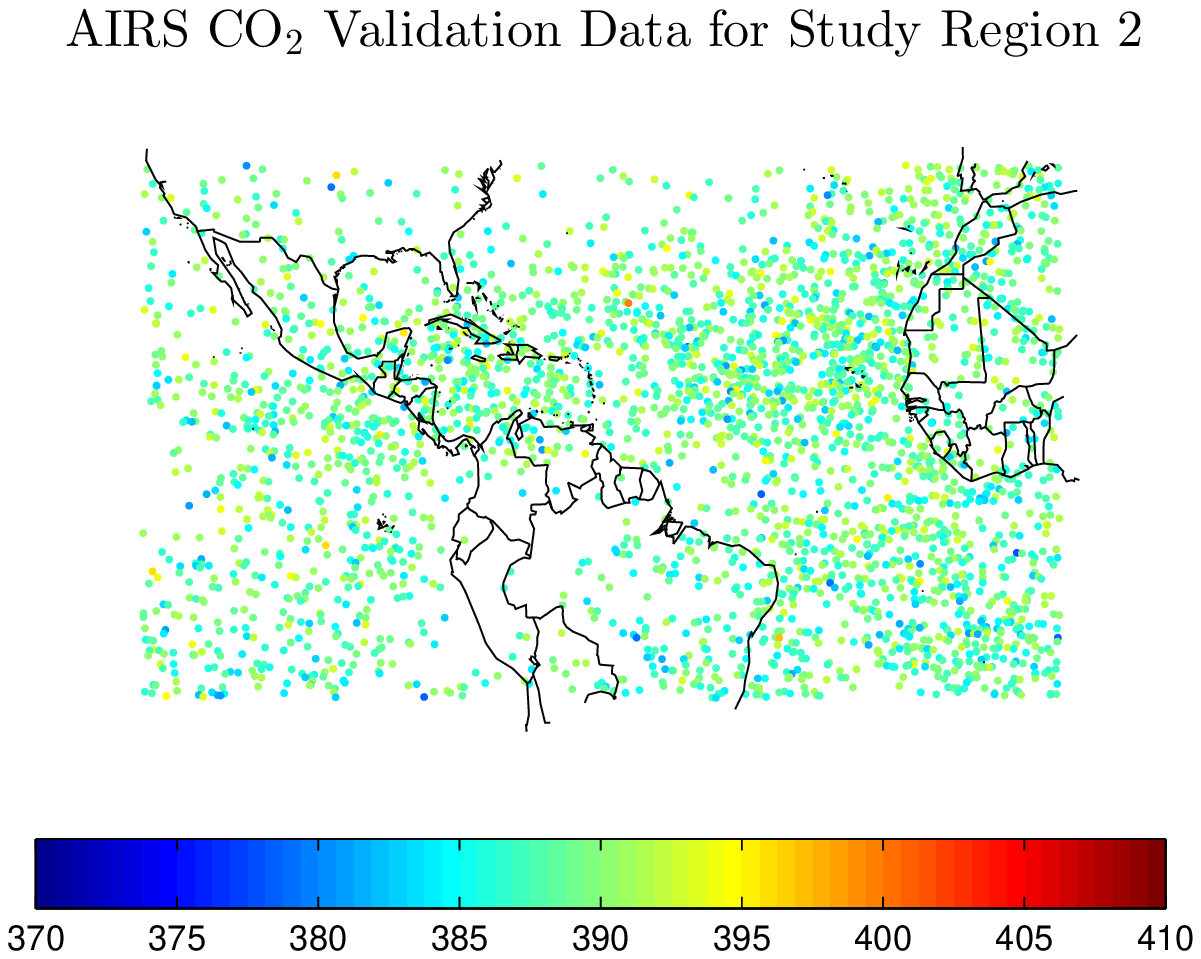}\\
\end{tabular}
\caption{A spatial dataset made up of 9 days of measurements of mid-tropospheric $\mathrm{CO}_{2}$ in parts per million (ppm). The data considered are between $-60^{\circ}$ degrees and $90^{\circ}$ degrees latitude from Februrary 1 through Februrary 9, 2010. The data are randomly split into observed and testing datasets with $n = 12,358$ and $m = 3,090$, respectively. }\label{fig:6}
\end{center}
\end{figure}

\clearpage
\begin{figure}[htp]
\begin{center}
\begin{tabular}{c}
\includegraphics[width=11.5cm,height=6cm]{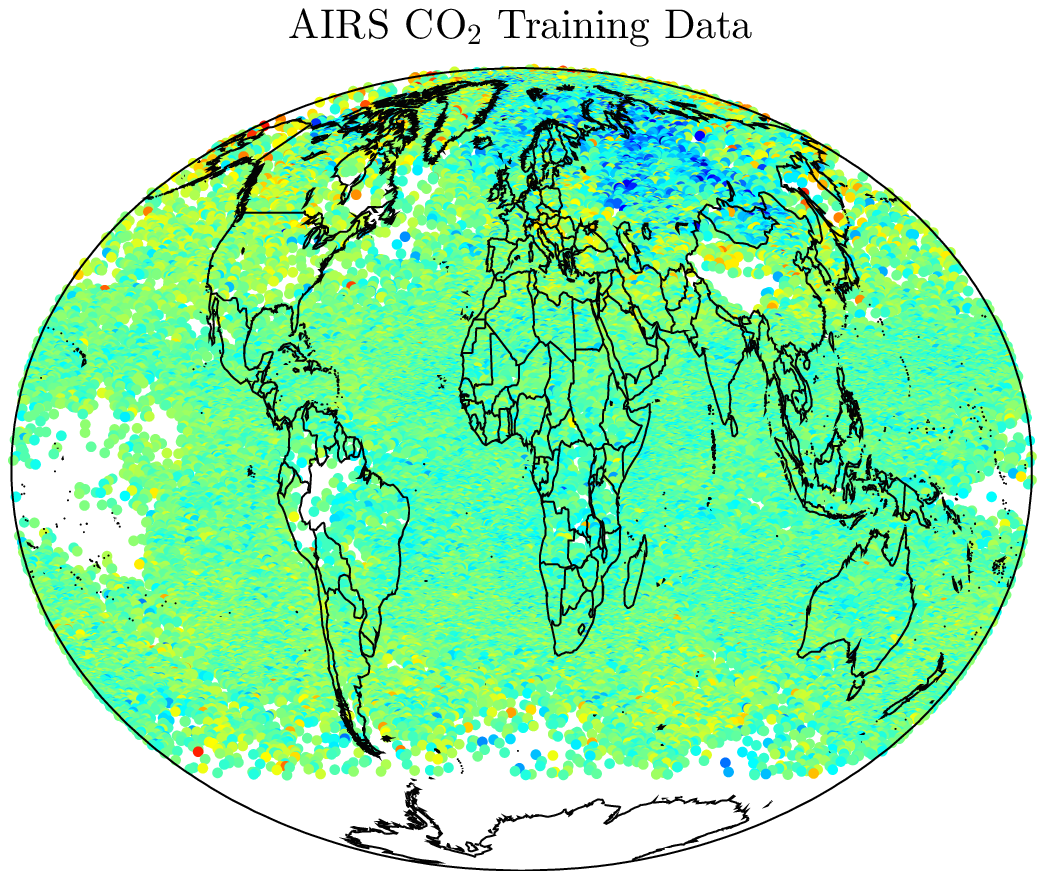}\\
\newline
\includegraphics[width=12cm,height=7.5cm]{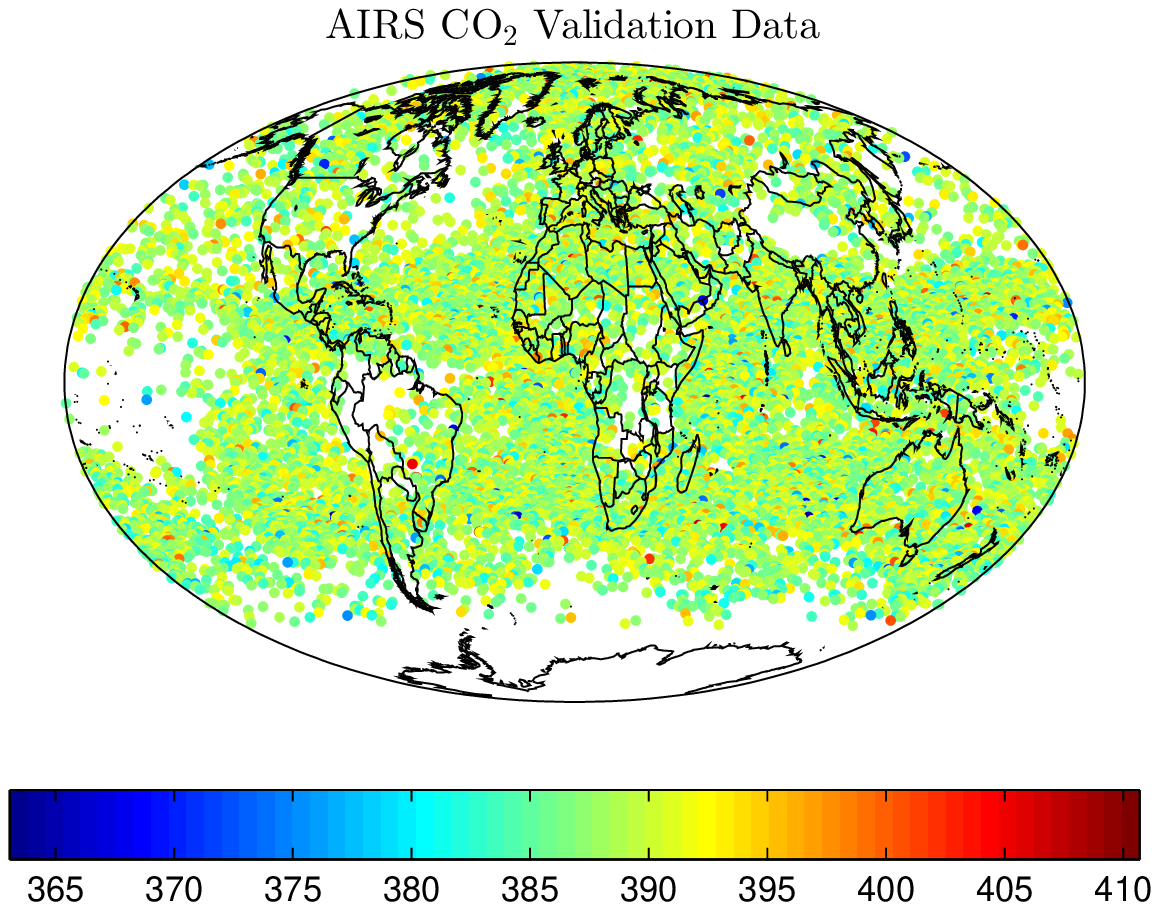}\\
\end{tabular}
\caption{A spatial dataset made up of 9 days of measurements of global mid-tropospheric $\mathrm{CO}_{2}$ in parts per million (ppm). The data considered are between $-60^{\circ}$ degrees and $90^{\circ}$ degrees latitude from Februrary 1 through Februrary 9, 2010. The data are randomly split into observed and testing datasets with $n = 44,621$ and $m = 2,000$, respectively. }\label{fig:6}
\end{center}
\end{figure}

\clearpage
\begin{figure}
\begin{tabular}{c}
\includegraphics[width=16cm,height=20cm]{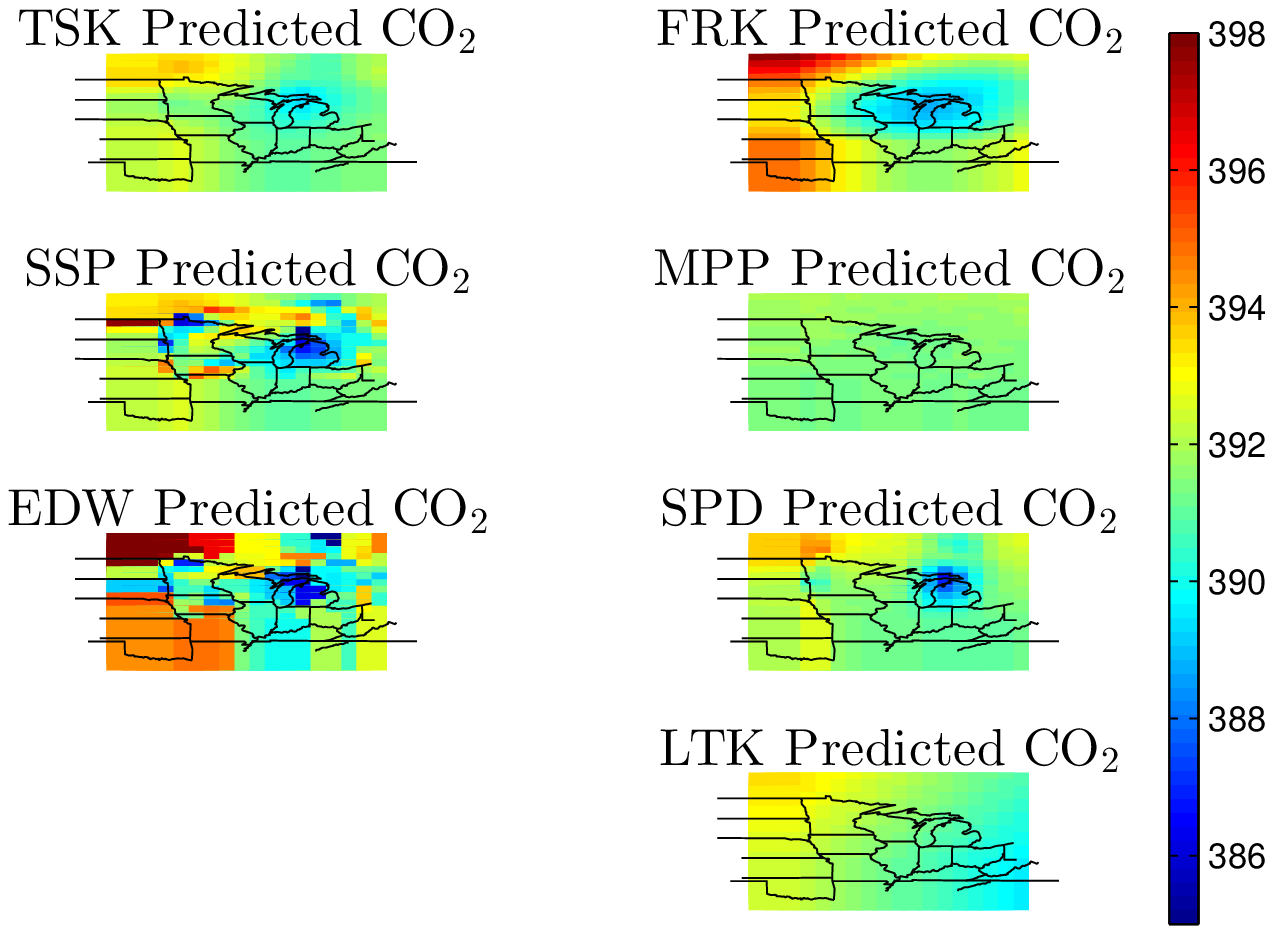}
\end{tabular}
\caption{Spatial prediction of mid-tropospheric $\mathrm{CO}_{2}$ concentrations using TSK, SSP, EDW, FRK, MPP, SPD, and LTK. Predictions are indicated in the title headings and are mapped over Study Region 1.}\label{fig:7}
\end{figure}

\clearpage
\begin{figure}
\centering
\begin{tabular}{c}
\includegraphics[width=16cm,height=20cm]{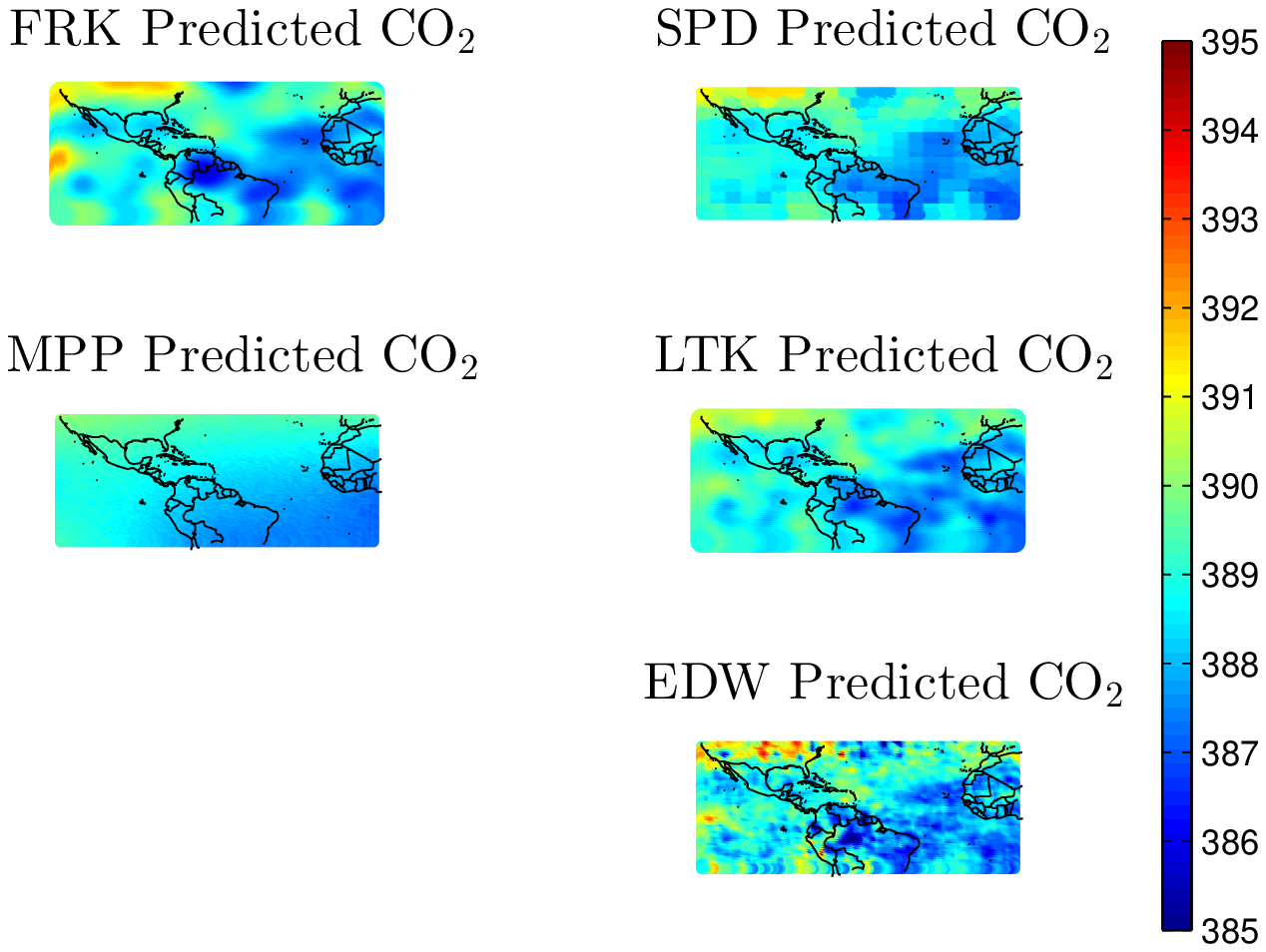}
\end{tabular}
\caption{Spatial prediction of mid-tropospheric $\mathrm{CO}_{2}$ concentrations using FRK, MPP, SPD, LTK, and EDW. Predictions are indicated in the title headings and are mapped over Study Region 2.}\label{fig:7}
\end{figure}

\clearpage
\begin{figure}
\centering
\begin{tabular}{c}
\includegraphics[width=16cm,height=9cm]{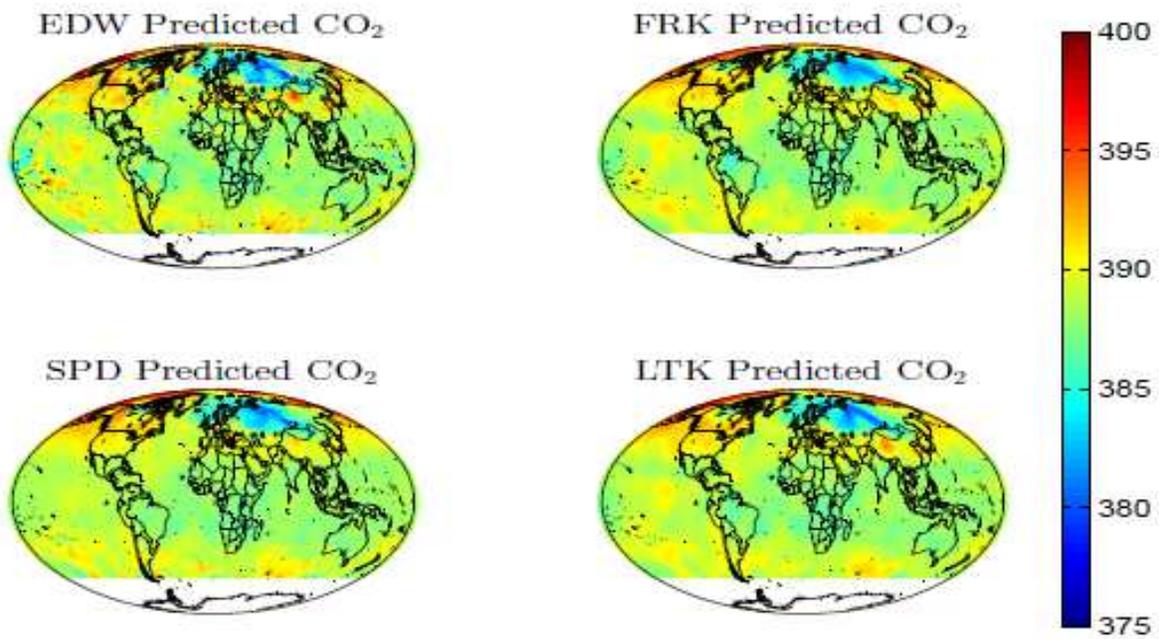}
\end{tabular}
\caption{Global spatial prediction of mid-tropospheric $\mathrm{CO}_{2}$ concentrations using EDW, SPD, FRK, and LTK. Predictions are indicated in the title headings and are mapped over Study Region 3. Note that there is no prediction given below latitude $-60^{\circ}$, since AIRS has not released any observations there.}\label{fig:7}
\end{figure}

\clearpage
\begin{table}[h!]
\caption{Results from Study Region 1 (Section 3.1) for the root average squared testing error (RSTE), CPU time, and peak memory-usage by predictor. These quantities are produced using the data in Figure 1.}\label{tab:3}
\begin{center}
\begin{tabular}{ |c|c|c|c|c|c|}
\hline
	Predictor &	RSTE &	PMCC &	\specialcell{Lag-1\\Semivariogram} & \specialcell{CPU Time\\(in minutes)}& \specialcell{Peak Memory\\Usage (in MB)}\\ \hline
	TSK & 4.7063 &  -0.4845 &	0.5739 & 0.20&  171.08\\ \hline
	SSP & 4.7151 &  N/A &	4.9746 &  0.02&  1,043.80\\ \hline
	EDW & 4.7703 &  N/A &	7.5466 &  0.30&  733.39\\ \hline
	FRK	&	4.3097 &  12.5612 &	2.1298 & 1.01 &  791.12\\ \hline
	MPP & 4.9084 &  -0.5873 & 0.0339 &  3.37&  239.51\\ \hline
	SPD	&	4.7399 &  26.2548 &	1.1271 & 0.24 &  143.14\\ \hline
	LTK	&	5.0163 &  39.5806 &	0.2536 & 2.73 &  205.84\\ \hline
\end{tabular}
\end{center}
\end{table}

\clearpage
\begin{table}[h!]
\caption{Results from Study Region 2 (Section 3.2) for the root average squared testing error (RSTE), CPU time, and peak memory-usage by predictor. These quantities are produced using the data in Figure 2.}\label{tab:3}
\begin{center}
\begin{tabular}{ |c|c|c|c|c|c|}
\hline
	Predictor &	RSTE &	PMCC &	\specialcell{Lag-1\\Semivariogram} & \specialcell{CPU Time\\(in minutes)}& \specialcell{Peak Memory\\Usage (in MB)}\\ \hline
	EDW & 3.0396 &  N/A & 0.6966 & 6.36&  850.0470\\ \hline
	FRK	&	3.0067 & 12.6155& 0.2075 & 0.52 &  841.0030\\ \hline
	MPP & 2.9243 & -1.0327 & 0.0164 & 216.79&  2042.6\\ \hline
	SPD	&	2.9630 & 70.0529 & 0.1243 & 0.47 &  111.18\\ \hline
	LTK	&	2.9855 & 27.3636 & 0.1470 & 1.72 &  1,971.8\\ \hline
\end{tabular}
\end{center}
\end{table}

\clearpage
\begin{table}[h!]
\caption{Results from Study Region 3 (Section 3.3) for the root average squared testing error (RSTE), CPU time, and peak memory-usage by predictor. These quantities are produced using the data in Figure 3.}\label{tab:3}
\begin{center}
\begin{tabular}{ |c|c|c|c|c|c|}
\hline
	Predictor &	RSTE &	PMCC &	\specialcell{Lag-1\\Semivariogram} & \specialcell{CPU Time\\(in minutes)}& \specialcell{Peak Memory\\Usage (in MB)}\\ \hline
	EDW & 4.0799 & N/A & 1.6088 & 90.68&  691.57\\ \hline
	FRK	&	3.9841 & 12.0974 & 0.5080 & 0.51 &  1,025.40\\ \hline
	SPD	&	3.9882 & 53.1760 & 2.1121 & 4.72 &  165.19\\ \hline
	LTK	&	4.0026 & 45.1762 & 0.1440 & 85.13 &  490.60\\ \hline
\end{tabular}
\end{center}
\end{table}

\end{document}